\documentclass[aps,prb,twocolumn,amssymb,float,epsfig]{revtex4}
\usepackage{epsfig,psfrag,subfigure,amsopn}
\usepackage{graphicx,psfrag,subfigure,float}
\usepackage{color}
\usepackage{amssymb,amsbsy,amsmath}

\newcommand\beq{\begin{equation}}
\newcommand\eeq{\end{equation}}
\newcommand\bea{\begin{eqnarray}}
\newcommand\eea{\end{eqnarray}}
\newcommand\al{\alpha}
\newcommand\be{\beta}

\newcommand\de{\delta}
\newcommand\ep{\epsilon}
\newcommand\De{\Delta}
\newcommand\si{\sigma}

\newcommand\la{\lambda}

\newcommand\ta{\theta}
\newcommand\dg{\dagger}
\newcommand\pa{\partial}
\newcommand\non{\nonumber}
\newcommand\noi{\noindent}

\newcommand\ig{\includegraphics}
\newcommand\bib{\bibitem}

\begin{document}

\title{Transport across a system with three $p$-wave superconducting wires: 
effects of Majorana modes and interactions}

\author{Oindrila Deb, Manisha Thakurathi and Diptiman Sen}
\affiliation{Centre for High Energy Physics, Indian Institute of Science, 
Bengaluru 560 012, India}

\date{\today}

\begin{abstract}
We study the effects of Majorana modes and interactions between electrons
on transport in a 
one-dimensional system with a junction of three $p$-wave superconductors (SCs)
which are connected to normal metal leads. For sufficiently long SCs, there 
are zero energy Majorana modes at the junctions between the SCs and the leads, 
and, depending on the signs of the $p$-wave pairings in the three SCs, there 
can also be one or three Majorana modes at the junction of the three SCs. 
We show that the various sub-gap conductances have peaks 
occurring at the energies of all these modes; we therefore get a rich pattern 
of conductance peaks. Next, we use a renormalization group approach 
to study the scattering matrix of the system at energies far from the SC gap. 
The fixed points of the renormalization group flows and their stabilities are 
studied; we find that the scattering matrix at the stable fixed point is 
highly symmetric even when the microscopic scattering matrix and the 
interaction strengths are not symmetric. We discuss the implications of this 
for the conductances. Finally we propose an experimental realization of this 
system which can produce different signs of the $p$-wave pairings in the 
different SCs.
\end{abstract}

\maketitle

\noi {\large {\bf 1 ~Introduction}}
\vspace*{.4cm}

The subject of Majorana fermions has been extensively studied in recent 
years, both from the fundamental physics point of view~\cite{beenakker} as well
as for possible applications in topological quantum computation~\cite{nayak}.
A well-studied example of a system which hosts Majorana modes is the Kitaev 
chain, which is a spin-polarized $p$-wave superconductor (SC) in one 
dimension. This system is known to have topological phases in which there is 
a zero energy Majorana mode at each end of a long chain~\cite{kitaev1}. This 
system and others similar to it have been theoretically studied in a number of
papers~\cite{beenakker,lutchyn1,fu,potter,shivamoggi,vish,fidkowski1,brouwer,
alicea1,fulga,stanescu1,ganga,stoud,fidkowski2,chung,sela,jiang,dirks,tewari1,
beri,gibertini,lobos,hutzen,pientka1,roy,cook,sticlet1,jose,klino1,alicea2,
hassler,sau1,liu,lang,niu,degottardi1,leijnse,nadj,degottardi2,cai,kesel,
klino2,adagideli,mano,lobos2,kashuba,ghaza,guigou,klino3,door,kao,hegde,
thakurathi,spanslatt,dasmandal}. Several experiments inspired by these models 
have looked for Majorana modes, either through a zero bias conductance 
peak~\cite{kouwenhoven,deng,das1,finck,lee,nadj2,pawlak} or through the 
fractional Josephson effect~\cite{rokhinson}. The zero bias peak occurs 
because the Majorana mode (which lies at zero energy for a long enough 
system) allows an electron to tunnel from a normal metal 
lead into the superconductor where it turns into a Cooper pair; this
gives rise to a perfect Andreev reflection back into the lead.

The Kitaev chain has been generalized in a number of ways; some of these 
generalizations give rise to more than one Majorana mode at each end of the 
SC region~\cite{fidkowski1,potter,beri,niu,degottardi2}. The case of additional
Majorana modes appearing inside the {\it bulk} of the SC region (rather than 
at the ends) is less well studied. Strictly speaking, the term Majorana 
modes is used to refer to states with exactly zero energy. In this paper, we 
will discuss various sub-gap modes which lie within the SC gap but may or 
may not lie at zero energy. All these modes originate from the zero energy 
Majorana modes which occur if the SC wires are very long, but they hybridize 
and thereby move away from zero energy when the wire lengths are finite.
It is interesting to study the effect of all such sub-gap modes on the 
electronic transport across the system. 

The effect of interactions between the electrons is also of interest.
In one dimension, interactions are known to have a dramatic effect, turning 
the system into a Tomonaga-Luttinger liquid. The fate of the Majorana modes 
in the presence of interactions has been studied extensively~\cite{ganga,sela,
lobos,hutzen,hassler,stoud,fidkowski2,mano,kashuba,ghaza}. We will examine a 
different aspect of our system, namely, the fixed points of the conductances 
which arise from the renormalization group flows induced by interactions. 

In this paper, we will study the conductances of a one-dimensional system 
consisting of a junction of three finite-sized $p$-wave superconductors, 
each of which is in turn connected to a semi-infinite normal metal (NM) 
lead. We will study transport in two different regimes of the energy, 
namely, inside the SC gap and far outside the gap.

In the first part of this paper, we will study the Majorana and related 
sub-gap modes and their effect on the sub-gap conductances. A motivation for 
studying a junction of three $p$-wave superconductors is that a topological
quantum computation using Majorana modes in quasi-one-dimensional systems would
involve braiding of two Majorana modes without the two modes ever coming close 
to each other. In order to achieve this we must consider junctions of three 
or more wires, so that a Majorana mode can be moved from the first wire to the 
second without disturbing a Majorana mode lying on the third wire. (A single 
superconducting wire does not allow braiding of Majorana modes). We will show 
that there are two interesting possibilities for a junction of three 
superconducting wires which have rather different properties: the $p$-wave 
pairing amplitude $\De$ can have the same sign in all three SCs, or $\De$ can 
have the same sign in two of the SCs and the opposite sign in the third SC. 
We will see that the number of zero energy Majorana modes is different in 
the two cases when the SC wires are very long (namely, much longer than the 
decay length which will be discussed below). 
In the first case, there are six Majorana modes (three at the NM-SC 
junctions, and three more at the junction of the three SCs), while in the 
second case, there are four Majorana modes (three at the NM-SC junctions and 
only one at the junction of the SCs). The fact that there can be three 
Majorana modes at a junction of three $p$-wave superconducting wires 
has not been pointed out earlier as far as we know.

We would like to note that many kinds of junctions have been 
studied before, 
such as junctions of three wires or three spin-1/2 chains (which can be mapped
to fermionic chains)~\cite{alicea1,zhou,tsvelik}, a superconducting junction 
of non-superconducting wires~\cite{erik,alt}, and a junction of multiple 
non-superconducting wires with a superconducting wire~\cite{affleck}. In 
this paper, we will not consider the effect of interactions at the junction 
(i.e., charging energy)
which leads to the Kondo-Majorana physics studied in Refs.~\onlinecite{erik} 
and \onlinecite{alt}. However we will discuss the effect of interactions in 
the wires away from the junction as mentioned below.

Electronic transport across a NM-SC-NM system has been studied for many 
years~\cite{blonder,kasta,ks01,yoko,hayat}. The presence of a SC means that 
there will be both normal reflection and transmission and Andreev reflection 
and transmission~\cite{blonder,andreev}. Hence there are two kinds of 
differential conductances which can be measured in this NSN system: 
a conductance from one NM lead to the other NM lead which we will call 
$G_N$, and a Cooper pair conductance from a NM lead to the SC which we 
will call $G_C$. In our system with three SCs and three NM leads, we will
consider an electron incident from one of the leads (called NM1); we can
then have a conductance $G_C$ from that lead to the SC, and conductances
$G_{N2}$ and $G_{N3}$ from that lead to the other two leads called NM2 and
NM3 respectively. We will use a continuum model for this system, rather
than the Kitaev model which is defined on a lattice. We will first present
the boundary conditions at the three NM-SC junctions which follow from 
the conservation of both the probability and charge currents. We will then 
discuss the boundary conditions at the junction of the three SCs; this will 
turn out to involve a $3 \times 3$ Hermitian matrix $\bf M$ which determines 
how a current incident on the junction from one of the SCs either gets 
reflected back to that SC or gets transmitted to the other two SCs. Using all 
these boundary conditions, we will numerically calculate $G_C, ~G_{N2}$ and 
$G_{N3}$ as functions of the energy $E$ of the electron incident from the 
lead NM1 and the lengths $L_j$ of the three SCs. We will see that $G_C$ has 
a rich structure of peaks when $E$ lies in the superconducting gap. The 
conductance calculation will be followed by the discussion 
of a box made of only the SCs with hard wall boundary conditions (namely,
with no NM leads). We will numerically calculate the energies of the 
sub-gap modes in this system and show that this explains the locations
of the peaks in the conductance $G_C$ of the system with NM leads. 

In the second part of the paper, we will study the effect of interactions 
between the electrons on the conductances of the system at energies which are
far from the SC gap, namely, when $|E| \gg |\De|$. 
(The Majorana and other sub-gap modes play no role in this part; this is 
therefore complementary to the first part of the paper 
where we look at the sub-gap conductances).
It is common to use the technique of bosonization to study one-dimensional
systems with interacting electrons~\cite{gogolin}. However, for reasons that 
will be discussed below, it turns out to be difficult to use bosonization 
when there is a junction of three or more wires and superconductivity is 
present. (Three-wire junctions without superconductivity have been studied in 
Refs.~\onlinecite{nayak2,trau,chamon,bella,agar} using bosonization and in
Ref.~\onlinecite{meden} using functional renormalization group methods.
Junctions between a superconducting wire and multiple non-superconducting
wires with interactions have been studied using bosonization~\cite{affleck}).
We will therefore use a different method which is valid when the strength
of the interactions is weak~\cite{yue,lal,das2,saha}. Using this method we 
will find that the scattering matrix $S$ which characterizes the junction 
of three wires effectively becomes a function of the length scale, and we will 
find a renormalization group (RG) equation for $S$. Using the RG equation, 
we will study how $S$ varies with the length scale; we will find the
fixed points of the RG equation and study their stabilities. We will then 
discuss the implications of this for the conductances of the system when
the wire lengths are large or the temperature is low.

The plan of the paper is as follows. In Sect.~2, we introduce 
the model for the NSN system and derive the boundary conditions at 
the junctions between the SCs and the NM leads and at the junction of
three SCs. We show how this can be used to derive the various differential 
conductances $G_{Nj}$ and $G_C$ at energies lying inside the SC gap. 
In Sect.~3, we numerically calculate $G_{Nj}$ and $G_C$ for two 
cases: when the $p$-wave pairings $\De_j$ have the same sign in all the 
three wires and when one of them has a different sign from the other two. 
We discuss three different regimes of the lengths of the SC wires: these 
lengths can be less than, a little larger than, and much larger
than the decay length of the sub-gap modes which appear near the different
junctions. The locations of the peaks in the conductances are quite different
in the three length regimes and also in the two cases of the relative signs 
of the $\De_j$. To understand these differences, we have numerically found
the energies of the sub-gap modes and shown that they precisely match the
locations of the conductance peaks. In Sect.~4, we provide
analytical arguments to show that the number of zero energy Majorana modes 
at the junction of three long $p$-wave SCs is three if the $\De_j$'s have 
the same sign but is one
if one of the $\De_j$'s has a different sign. In Sect.~5, we study 
the effect of interactions between the electrons on the conductance at 
energies far from the SC gap (but much smaller than the band width of the
system). To this end we consider
a junction of three NM wires which meet in a SC region, and we derive the
RG equations for the scattering matrix of this system for the case where
the electrons interact weakly with each other. In Sect.~6, we study
the fixed points and stabilities of the RG equations and show that this
can lead to power-law dependences of the conductances on the wire lengths or
temperature. In Sect.~7 we show how a system of three $p$-wave
SCs can be experimentally realized and how the two cases of all the $\De_j$'s
having the same sign or one of them having a different sign can be fabricated.
We end in Sect.~8 with a summary of our results. 

\vspace*{.6cm}
\noi {\large {\bf 2 ~Model for a system of three SC wires making a 
$Y$-junction}}
\vspace*{.4cm}

\begin{figure}[h]
\begin{center} \ig[width=3.4in]{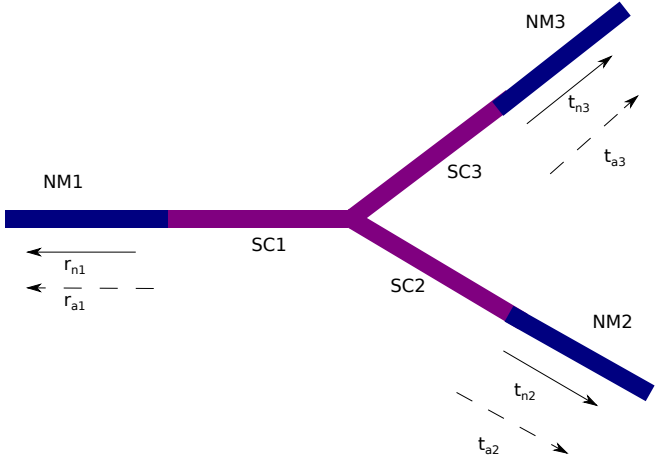}
\caption{Schematic picture of a $Y$-junction of three wires
labeled as 1-3. The inner (lighter) regions of the wires are $p$-wave SCs 
while the outer (darker) regions are NM leads. For an electron incident from 
NM1, six amplitudes are shown in the leads: $r_{n1}, ~r_{a1}$ are normal and 
Andreev reflections in NM1, $t_{n2}, ~t_{a2}$ are normal and Andreev 
transmissions in NM2, and $t_{n3}, ~t_{a3}$ are normal and Andreev 
transmissions in NM3.} \label{fig:3sc1} \end{center} \end{figure}

We begin with a continuum model for a $Y$-junction of three $p$-wave SC 
wires in one dimension as shown in Fig.~\ref{fig:3sc1}. A NM lead is attached 
to the end of each SC where there is a barrier modeled by a $\de$-function 
potential. Each NM-SC system has a coordinate system $x_j$, with $j=1,2,3$; we 
define the point where the three SCs meet as the origin $x_j = 0$ for all $j$.
As we move away from this junction, $x_j$ will be taken to increase. The 
different SCs may have different lengths and lie in the regions $0 \le x_j 
\le L_j$; the leads lie in the region $L_j \le x_j < \infty$. 

Let us denote the wave function in wire $j$ as $\psi= (c,~ d)^T$, where 
$c(x_j,t), ~d(x_j,t)$ are the electron and hole components respectively 
(we assume that the electrons are spin-polarized and will therefore 
ignore the spin label). The Hamiltonian in each wire can be written as
\bea H &=& \int_0^\infty dx_j~[ c^\dg (- \frac{\hbar^2 \pa_{x_j}^2}{2m} -\mu) 
c ~-~ d^\dg (- \frac{\hbar^2 \pa_{x_j}^2}{2m} - \mu ) d \non \\
&& ~~~~~~~~~- \frac{i\De_j}{k_F} ~( c^\dg \pa_{x_j} d + d^\dg \pa_{x_j} c ) ], 
\label{Ham} \eea
where $\mu$ is the chemical potential, $k_F = \sqrt{2 m \mu}/\hbar$ is the 
Fermi wave number, and $\De_j$ is the $p$-wave superconducting pairing 
amplitude which will be assumed to be real everywhere; we will set $\De_j=0$ 
in the NM leads. (We will generally set $\hbar = 1$ in this paper, except in 
places where it is required for clarity). The Heisenberg equations of motion 
$i \pa_t c = - [H,c]$ and $i \pa_t d = - [H,d]$ imply that
\bea i \pa_t c &=& -\left( \frac{\pa_{x_j}^2}{2m} +\mu \right)~c ~-~
\frac{i\De_j}{k_F} ~\pa_{x_j} d, \non \\
i \pa_t d &=& \left( \frac{\pa_{x_j}^2}{2m} +\mu \right)~d ~-~ 
\frac{i\De_j}{k_F} ~\pa_{x_j} c. \label{eom} \eea
For a wave function which varies in space as $e^{\pm ik x_j}$, the energy is 
given by $\pm [k^2/(2m) - \mu]$ if $\De_j = 0$, and by $\pm \sqrt{[k^2/(2m) - 
\mu]^2 + \De_j^2 (k/k_F)^2}$ if $\De_j \ne 0$. The corresponding wave functions 
will be presented below. We see that the energy spectrum in the $j$-th SC has 
a gap equal to $2\De_j$ at $k= \pm k_F$.

Let us define the particle density $\rho_p=c^\dg c +d^\dg d$ (this counts
electrons and holes with the same sign) and the charge density $\rho_c= 
c^\dg c -d^\dg d$ (which counts electrons and holes with opposite signs; we 
are ignoring a factor of electron charge here). Using Eqs.~\eqref{eom} and 
the equations of continuity $\pa_t {\rho_p} + \pa_x J_p = 0$ and $\pa_t 
{\rho_c} + \pa_x J_c = 0$, we find the particle and charge currents to 
be~\cite{blonder,soori}
\bea J_p &=& \frac{i}{2m} ~[-~c^\dg\pa_x c ~+~ \pa_x c^\dg c ~+~ d^\dg \pa_x 
d ~-~ \pa_x d^\dg d] \non \\
&& + ~\frac{\De}{k_F} ~(c^\dg d + d^\dg c), \non \\
J_c &=& J_1 ~+~ \int_0^x dx' ~J_2 (x'), \non \\
J_1 &=& \frac{i}{2m} ~[- ~c^\dg \pa_x c ~+~ \pa_x c^\dg c ~-~ d^\dg \pa_x d ~+~
\pa_x d^\dg d], \non \\
J_2 &=& - ~\frac{2\De}{k_F} ~(\pa_x c^\dg d + d^\dg \pa_x c). \label{curr} \eea
The last term, $J_2$, can be interpreted as the contribution of Cooper pairs 
to the charge current; note that it vanishes in the NM where $\De = 0$.

The boundary conditions at the NM-SC junctions at $x_j=L_j$ can be found 
by demanding that the currents $J_p$ and $J_c$ be conserved at those points. 
At the junction $x_1=L_1$, let us consider the wave functions 
$\psi_{nm1} =(c_{nm1},~d_{nm1})^T$ and $\psi_{sc1}= (c_{sc1}, ~d_{sc1})^T$ at 
the points $x_1=L_1+\ep$ and $x_1=L_1-\ep$, i.e., in the NM and SC regions 
respectively. The condition $J_p(L_1+\ep)= J_p (L_1-\ep)$ implies that 
\bea && \frac{i}{2m}~[- ~c_{nm1}^\dg\pa_{x_1} c_{nm1} ~+~ \pa_{x_1} 
c_{nm1}^\dg c_{nm1} \non \\ 
&& ~~~~~~+~ d_{nm1}^\dg \pa_{x_1} d_{nm1} ~-~ \pa_{x_1} d_{nm1}^\dg 
d_{nm1}] \non \\
&=& \frac{i}{2m} ~[-~ c_{sc1}^\dg \pa_{x_1} c_{sc1} ~+~ \pa_{x_1} c_{sc1}^\dg 
c_{sc1} \non \\
&& ~~~~~~+~d_{sc1}^\dg \pa_{x_1} d_{sc1} ~-~ \pa_{x_1} d_{sc1}^\dg d_{sc1}] 
\non \\
&& ~+ ~\frac{\De_1}{k_F} ~(c_{sc1}^\dg d_{sc1} ~+~ d_{sc1}^\dg c_{sc1}). \eea
The simplest way of satisfying this condition is to set
\bea c_{sc1}&=& c_{nm1}, \non \\
d_{sc1}&=&d_{nm1}, \non \\
\pa_{x_1} c_{sc1}~+~ \frac{i\De_1}{v_F} ~d_{sc1} &=& \pa_{x_1} c_{nm1}, \non \\
\pa_{x_1} d_{sc1}~-~ \frac{i\De_1}{v_F} ~c_{sc1} &=& \pa_{x_1} d_{nm1}, 
\label{bc1} \eea
where $v_F = k_F/m$ is the Fermi velocity.
The first two equations above mean that the wave function is continuous
while the last two equations imply that the first derivative is discontinuous
in a particular way. We now find that Eqs.~\eqref{bc1} imply that the
charge current is conserved, namely, $J_c(L_1-\ep)= J_c(L_1+\ep)$. 
Next, let us consider what happens if a $\de$-function potential of strength 
$\la$ is present at the junction at $x_1=L_1$; note that the dimension 
of $\la$ is energy times length. (This potential is physically
motivated by the fact that in many experiments, the NM leads are weakly 
coupled, by a tunnel barrier, to the SC. This can be modeled by placing a 
$\de$-function potential with a large strength at the junction).
Now there will be an additional discontinuity in the first derivative
at $x_1=L_1$; this is found by integrating over the $\de$-function which gives
\beq \pa_{x_1} \psi_{nm1} (L_1+\ep) ~-~ \pa_{x_1} \psi_{sc1} (L_1-\ep) ~=~ 2m 
\la ~\psi_{nm1}(0). \eeq
Hence Eqs.~\eqref{bc1} must be modified to
\bea c_{sc1}&=&c_{nm1} , \non \\
d_{sc1}&=&d_{nm1}, \non \\
\pa_{x_1} c_{sc1}~+~ \frac{i\De_1}{v_F} ~d_{sc1}~+~2m \la ~c_{sc1} &=& 
\pa_{x_1} c_{nm1}, \non \\ 
\pa_{x_1} d_{sc1}~-~ \frac{i\De_1}{v_F} ~c_{sc1}~+~2m \la ~d_{sc1} &=& 
\pa_{x_1} d_{nm1}. \label{bc2} \eea
For the other two NM-SC junctions at $x_2=L_2$ and $x_3=L_3$, we will get 
boundary conditions similar to Eqs.~\eqref{bc2}.

We have found the boundary conditions at the NM-SC junctions. Now we look at 
the junction $x_j = 0$ where the three SCs meet. We can find the boundary 
conditions at this junction using the conservation of the particle 
current $J_p$; this implies that 
\bea \sum_j {J_{p_j}} &=& \frac{i}{2m} ~\sum_j {[-~c_{scj}^\dg\pa_{x_j}
c_{scj}~+~ \pa_{x_j} c_{scj}^\dg c_{scj} ]} \non \\ 
&& +~\frac{i}{2m}~ \sum_j {[d_{scj}^\dg \pa_{x_j} d_{scj} ~-~ \pa_{x_j} 
d_{scj}^\dg d_{scj}]} \non \\
&& +~\sum_j {\frac{\De_j}{k_F} ~(c_{scj}^\dg d_{scj} + d_{scj}^\dg c_{scj})} 
\non \\
&=& 0. \label{bc4} \eea
The simplest way of satisfying this condition is to set 
\bea \pa_x c_{sc} &=& {\bf M} \cdot c_{sc} ~+~ {\bf N} \cdot d_{sc}, \non \\
\pa_x d_{sc} &=& {\bf M^*} \cdot d_{sc} ~+~ {\bf N^*} \cdot c_{sc}, 
\label{bc5} \eea
where $c_{sc} \equiv (c_{sc1},c_{sc2}, c_{sc3} )^T$, $d_{sc} \equiv 
(d_{sc1},d_{sc2}, d_{sc3})^T$, and ${\bf M}, ~{\bf N}$ are $3 \times 3$ 
matrices. Substituting the above conditions in Eq.~\eqref{bc4}, we find 
that we must have 
\bea {\bf M}^\dg &=& {\bf M}, \non \\
{\bf N} ~+~ {\bf N^T} &=& -\frac{i2m}{k_F} ~{\bf \De}, \label{bc6} \eea 
where ${\bf \De}= \left( \begin{array}{ccc} \De_1 & 0 & 0 \\ 
0 & \De_2 & 0 \\ 
0 & 0 & \De_3 \end{array}\right)$. In our calculations we will assume that 
the magnitudes of all the $\De_j$'s are the same but the signs of $\De_j$ may 
vary with $j$. Now we will use the conservation of the charge current $J_c$
at the junction $x_j = 0$. The second term in the expression in 
Eq.~\eqref{curr} for the charge current, namely, $\int_0^{x_j} dx' ~J_2 (x')$ 
goes to zero as $x_j \to 0$. So the conservation of $J_c$ implies that 
\bea \sum_j {J_{c_j}} &=& \frac{i}{2m} ~\sum_j {[-~c_{scj}^\dg\pa_{x_j}
c_{scj}~+~ \pa_{x_j} c_{scj}^\dg c_{scj} ]} \non \\ 
&& ~+~ \frac{i}{2m} ~\sum_j {[-d_{scj}^\dg \pa_{x_j} d_{scj} + \pa_{x_j} 
d_{scj}^\dg d_{scj}]} \non \\ 
&=& 0. \label{bc7} \eea
Using Eq.~\eqref{bc5} in the above expression we find that
\beq {\bf N} ~=~ - ~\frac{im}{k_F} ~{\bf \De}. \label{bc8} \eeq
Hence Eq.~\eqref{bc5} becomes
\bea \pa_x c_{sc} &=& {\bf M} \cdot c_{sc} ~-~ \frac{im}{k_F}~ {\bf \De} 
\cdot d_{sc}, \non \\
\pa_x d_{sc} &=& {\bf M^*} \cdot d_{sc} ~+~ \frac{im}{k_F}~ {\bf \De} \cdot 
c_{sc}. \label{bc9} \eea

It is useful to note some symmetries of our system. 

\noi (i) Eqs.~\eqref{eom} are symmetric under time reversal (which changes 
$t \to - t$ and complex conjugates all numbers) if we 
transform $c \to c^*$ and $d \to - d^*$. This will also be a symmetry of 
Eqs.~\eqref{bc9} if ${\bf M^*} = {\bf M}$. Eq.~\eqref{bc6} then implies that 
${\bf M}$ must be both real and symmetric. We will assume this henceforth.

\noi (ii) Eqs.~\eqref{eom}, \eqref{bc2} and \eqref{bc9} remain invariant
under the transformation
\bea c(x_j) &\to& i ~c(x_j), ~~~~ d(x_j) ~\to~ -i ~d(x_j), \non \\
\De_j &\to& - ~\De_j, \label{dej} \eea
for all value of $j$ and $x_j$. This implies that the conductances
discussed below remain invariant if $\De_j \to - \De_j$ on all the wires.

We now use the boundary conditions discussed above to find the various
reflection and transmission amplitudes when an electron is incident from,
say, the NM1 lead with unit amplitude. In the presence of the SCs, 
the various scattering processes that may occur are as follows~\cite{blonder}.

\noi (i) an electron can be reflected back to the NM1 lead with amplitude
$r_{n1}$.

\noi (ii) a hole can be reflected back to the NM1 lead with amplitude $r_{a1}$. 
Charge conservation then implies that a Cooper pair must be produced
inside the region SC1.

\noi (iii) an electron can be transmitted to the NM2 lead with amplitude 
$t_{n2}$.

\noi (iv) a hole can be transmitted to the NM2 lead with amplitude $t_{a2}$. 
(This is usually called crossed Andreev reflection~\cite{law2}). Then 
charge conservation implies that a Cooper pair must be produced inside SC2.

\noi (v) an electron can be transmitted to the NM3 lead with amplitude 
$t_{n3}$.

\noi (vi) a hole can be transmitted to the NM3 lead with amplitude $t_{a3}$ 
which implies that a Cooper pair must be produced inside SC3.

If the energy $E$ of the electron (incident from the lead NM1) lies in the 
superconducting gap, i.e., $-\De \le E \le \De$ ($E$ can be interpreted as the 
bias between the chemical potentials of NM1 and the SCs), Eqs.~\eqref{eom} 
imply that the wave functions in the different regions must be of the form
\bea \psi_{nm1} &=& e^{-i(k_F+k) x_1} ~\left( \begin{array}{c}
1 \\
0 \end{array} \right) +~ r_{n1} ~e^{i(k_F+k) x_1} ~\left(\begin{array}{c}
1 \\
0 \end{array} \right) \non \\
&& + ~r_{a1} ~e^{i(-k_F+k) x_1} ~\left(\begin{array}{c}
0 \\
1 \end{array} \right), \non \\
\psi_{nm2} &=& t_{n2} ~e^{i(k_F+k) x_2} ~\left(\begin{array}{c}
1 \\
0 \end{array} \right) +~ t_{a2} ~e^{i(-k_F+k) x_2} ~\left(\begin{array}{c}
0 \\
1 \end{array} \right), \non \\
\psi_{nm3} &=& t_{n3} ~e^{i(k_F+k) x_3} ~\left(\begin{array}{c}
1 \\
0 \end{array} \right) +~ t_{a3} ~e^{i(-k_F+k) x_3} ~\left(\begin{array}{c}
0 \\
1 \end{array} \right), \non \\
\psi_{scj} &=& t_{1j} ~e^{ik_1 x_j} ~\left( \begin{array}{c}
1 \\
sgn(\De_j) ~e^{i\phi} \end{array} \right) \non \\
&& + ~t_{2j} ~e^{-ik_2 x_j} ~\left(\begin{array}{c}
1 \\
-sgn(\De_j) ~e^{-i\phi} \end{array} \right) \non \\
&& + ~t_{3j} ~e^{ik_3 x_j} ~\left(\begin{array}{c}
1 \\
sgn(\De_j) ~e^{-i\phi} \end{array}\right) \non \\
&& + ~t_{4j} ~e^{-ik_4 x_j} ~\left(\begin{array}{c}
1 \\
-sgn(\De_j) ~e^{i\phi} \end{array} \right), 
\label{wavefn1} \eea
where $e^{i\phi}= (E -i\sqrt{\De^2 -E^2})/\De$, and $sgn$ denotes the
sign function. (We have assumed that the 
magnitudes of the $\De_j$ are the same; hence $e^{i\phi}$ is the same for the 
three SCs). The top and bottom entries in the wave functions denote the 
particle and hole components. The wave functions in the NM leads are 
proportional to $e^{i(\pm k_F+k) x_j}$, where $k \ll k_F$, namely,
we are working close to the Fermi energy. In each SC, we have four modes;
two of these decay exponentially while the other two grow as we move away 
from the junction of the three SCs. We denote the wave numbers of these 
modes by $k_1, ~-k_2, ~k_3$ and $-k_4$. Defining the decay length 
\beq \xi ~=~ \frac{v_F}{\De \sqrt{1-(E/\De)^2}}, \label{xi} \eeq
we find that the decaying modes have 
\bea k_1 &=& k_F+i/\xi, \non \\ 
-k_2 &=& -k_F+i/\xi, \label{decay} \eea 
while the growing modes have 
\bea k_3 &=& k_F-i/\xi, \non \\
\ -k_4 &=& -k_F-i/\xi. \label{grow} \eea

{}From Eqs.~\eqref{bc2} we get the relation between the wave functions on 
the NM and SC sides at each of the NM-SC junctions. Similarly, from 
Eqs.~\eqref{bc6} and \eqref{bc8} we get the relation among the wave 
functions of the three SCs at the junction where they meet. We thus have 
eighteen equations for the eighteen unknowns $r_{n1}, ~r_{a1}, ~t_{n2}, ~
t_{a2}, ~t_{n3}, ~t_{a3}, ~t_{1j}, ~t_{2j}, ~t_{3j}$ and $t_{4j}$, where 
$j=1,2,3$. After solving these equations we can calculate the reflection 
and transmission probabilities. The conservation law for the probability 
current implies that
\beq |r_{n1}|^2 ~+~ |r_{a1}|^2 ~+~ |t_{n2}|^2 ~+~ |t_{a2}|^2 ~+~ |t_{n3}|^2 ~
+~ |t_{a3}|^2~=~ 1. \label{prob} \eeq 

The net probabilities for an electron to be transmitted from the NM1 lead 
to the NM2 and NM3 leads gives the differential conductances $G_{N2}$ and 
$G_{N3}$ respectively, where
\bea G_{N2} &=& |t_{n2}|^2 ~-~|t_{a2}|^2, \non \\
G_{N3} &=& |t_{n3}|^2 ~-~|t_{a3}|^2. \label{gn} \eea
The net probability for the electron to be reflected back to the NM1 lead is
\beq G_B ~=~ |r_{n1}|^2 ~-~ |r_{a1}|^2. \label{gb} \eeq
The remainder, denoted by the differential conductance $G_C$, is the 
probability for the electron to be transmitted into the SCs in the form of 
Cooper pairs. The conservation of charge current implies that
\bea G_C &=& 1 ~-~ G_{N2}~-~ G_{N3} ~-~ G_B \non \\
&=& 2 ~(|r_{a1}|^2 ~+~ |t_{a2}|^2 ~+~ |t_{a3}|^2), \label{gc} \eea
where we have used Eqs.~(\ref{prob}-\ref{gb}) to derive the last line in
Eq.~\eqref{gc}. [Actually, the differential conductances into the NM leads
2 and 3 and into the SCs are given by $e^2/(2\pi \hbar)$ times $G_{N2}$, 
$G_{N3}$ and $G_C$ respectively, where $e$ is the charge of an electron. 
However, we will ignore the factors of $e^2/(2\pi \hbar)$ in this paper 
and simply refer to $G_{Nj}$ and $G_C$ as the differential conductances.] 

We note that a differential conductance denotes $G = dI/dV$. To measure 
$G_{N2}$, $G_{N3}$ and $G_C$ in our system, we have to assume that there is a 
voltage bias $V$ between the NM1 lead on the one hand and the SCs and
the NM leads on the other (the SCs, NM2 and NM3 are taken to be at
the same potential). Namely, we choose the mid-gap energy in the SCs as zero,
and the Fermi energies in the NM1 lead as $E=eV$ and in NM2 and NM3 as zero. 
The differential conductances $G_{Nj}=dI_{Nj}/dV$ and $G_C = dI_C/dV$
are then the derivatives with respect to $V$ of the currents measured in the 
NM leads and in the SCs.

\vspace*{.6cm}
\noi {\large {\bf 3 ~Numerical results}}
\vspace*{.4cm}

\noi {\bf 3.1 ~All $\De_j$'s with the same sign}
\vspace*{.4cm}

In this section we present numerical results for $G_C$ and $G_{N2}$ as 
functions of the length $L_1$ of SC1 and the ratio $E/\De$ lying in the range 
$[-1,1]$. The length scale associated with the SC gap is $\eta = v_F/\De= 
k_F/{m\De}$. (This is different from the length $\xi$ introduced in 
Eq.~\eqref{xi} which depends on the energy $E$. Note that $\xi = \eta$ if 
$E=0$). We study three cases, namely, $L<\eta$, $L\sim \eta$ and $L> \eta$.

The values of the parameters that we have used to numerically calculate the 
conductances are as follows: $k_F=1$, $m=0.5$, $\la =5$, and $\De_1 = \De_2
= \De_3 = \De =0.1$. Throughout this section and in the next, we 
will take the matrix at the junction of the three SC wires to be of a form 
which is completely symmetric under any permutation of the three wires, \\
${\bf M} =\left ( \begin{array}{ccc} 1 & 1 & 1 \\ 
1 & 1 & 1 \\ 
1 & 1 & 1 \end{array}
\right)$. \\ 
The diagonal terms of ${\bf M}$ connect $\pa_x c ~(\pa_x d)$ and 
$c ~(d)$ in the same wire, while the off-diagonal terms connect $\pa_x c ~(
\pa_x d)$ and $c ~(d)$ in different wires. If we choose the off-diagonal 
and diagonal elements of ${\bf M}$ to be different, some of the results for 
the conductances may differ from what we will discuss below.

In the first row of Fig.~\ref{fig:3sc2}, we have shown top views of $G_C$. 
In the third row we show surface plots of $G_{N2}$ which is the 
conductance in NM2. In our calculations we have chosen $L_2$ and $L_3$ 
not to differ much from each other, so that $G_{N2}$ and $G_{N3}$ are 
similar. Hence we have presented $G_{N2}$ only. 
 
To understand the conductance plots better we consider another system 
which is made of three SC wires but is not connected to any NM leads; namely,
there is an infinite barrier at the ends of the SC wires. Thus this system is 
similar to a particle in a box but the box is made of three SC wires. 
The wave function is zero at the ends of the wires because of the infinite
barriers present there. At the junction of the three wires we use the same 
boundary condition that we derived in Eqs.~\eqref{bc9}. Using the boundary 
conditions we get a set of twelve linear homogeneous equations for the 
amplitudes $t_{1j}, ~t_{2j}, ~t_{3j}$ and $t_{4j}$ in the SC wires. These 
equations have a non-trivial solution if the determinant $D$ of the matrix 
constructed from these equations is zero. We plot the determinant $D$ as a 
function of $L_1$ and $E/\De$. The points where $D=0$ give the parameter 
values where sub-gap modes appear in the system. In the second row of 
Fig.~\ref{fig:3sc2}, we have shown the top views of $D$; 
the lightest regions correspond to the points closest to $D=0$. We expect 
that the positions of the peaks of $G_C$ will match the positions of the 
zeros of $D$ as $L_1$ and $E/\De$ are varied.

For the parameter values we have chosen, we find that $\eta=20$. In the 
first column of Fig.~\ref{fig:3sc2}, we take $L_1=3.5\pi$ to $7.5\pi$ (so that 
$L_1<\eta$), $L_2=6.7\pi$ and $L_3=6.3\pi$. In this regime we find that 
there are six sub-gap modes inside the SCs which is evident from the top
view of $G_C$ in Fig.~\ref{fig:3sc2} (a). If we look at a particular value 
of $L_1$ in that figure, we see six different modes at different energies. 
(Some of the modes look fainter than the others).
Three of the sub-gap modes lie near the NM-SC junctions as we know from 
earlier papers (see, for example, Ref.~\onlinecite{thakurathi}). The other 
three modes must lie near the three-wire junction. To explain the presence of 
modes at non-zero energies, we recall that the momenta in the SC have an 
imaginary part, giving rise to factors of $\pm x_j/\xi$ as we can see from the 
expression for $\psi_{scj}$ in Eq.~\eqref{wavefn1}. On wire $j$, the sub-gap 
wave functions of the form $e^{x_j/\xi}$ decay exponentially as we go away 
from the NM-SC junction, while the wave functions of the form $e^{-x_j/\xi}$ 
decay as we go away from the junction of three SC wires. When the lengths 
of the SCs are small compared to the decay length $\xi$ of the sub-gap modes, 
the amount of decay of the modes inside the SC will be small. Hence, the
sub-gap modes at the NM-SC junctions and at the junction of three SC wires 
will hybridize and their energies will split from zero. So the mixing of the 
sub-gap modes for SC wires with small lengths is responsible for the 
appearance of states at non-zero energies. These energies oscillate with 
$L_1$ and vanish at certain values of $L_1$.

As there are states inside the SCs and the lengths of the SCs are smaller
than the decay length $\xi$, there are two possibilities for an electron 
coming from the lead NM1.
 
\noi (i) It can enter the SC1 by coupling to the Majorana mode sitting there 
and then turn into a Cooper pair; a hole goes back into NM1 to conserve 
charge. In this process we get a finite $r_{a1}$ and hence a finite $G_C$.

\noi (ii) It can enter the SC1 similarly as described above and can then get 
transmitted to the other two NMs as the wave functions of the sub-gap modes 
decay very little inside the SCs. So we get finite transmission probabilities 
and hence a finite $G_{N2}$ and $G_{N3}$. From the figures in the first column
of Fig.~\ref{fig:3sc2}, it is clear that both $G_C$ and $G_{N2}$ are 
appreciable when $L_1<\eta$. 

In the third column of Fig.~\ref{fig:3sc2}, we choose $L_1=70.5\pi$ to 
$74.5\pi$, $L_2=68.3\pi$ and $L_3=68.7\pi$, so that all the lengths are much 
larger than the decay length $\xi$ of the sub-gap modes. In this regime 
the sub-gap modes are mostly localized at the ends of the SCs and at the 
junction of the three SC wires. So the coupling between the different sub-gap
modes are small, and their energies remain almost at zero. In 
Figs.~\ref{fig:3sc2} (c) and (i), we see that instead of six sub-gap modes at 
different energies, we now have all the sub-gap modes near zero energy 
(namely, they are almost Majorana modes). There is a small splitting in the 
energy around $E=0$ because of a small non-zero coupling between the different
modes. If we increase the various lengths more, we expect to see no splitting 
at all and all the modes should stay exactly at $E=0$. 

In the large length regime we find that $G_C$ almost approaches its highest 
value of 2 (in units of $e^2/h$) while $G_{N2}$ is very small. The reason 
for this is that the sub-gap modes are now almost decoupled from each other;
due to the exponential decay of their wave functions, the probability of an 
electron to travel through the SCs and get transmitted to the two NMs on the 
other side is very small. So the electron mostly transmits into the 
SCs and turns into a Cooper pair inside the SC1, so that $|r_a|^2 \simeq 1$. 

\begin{widetext}
\begin{center} \begin{figure}[H]
\subfigure[]{\ig[width=2.0in]{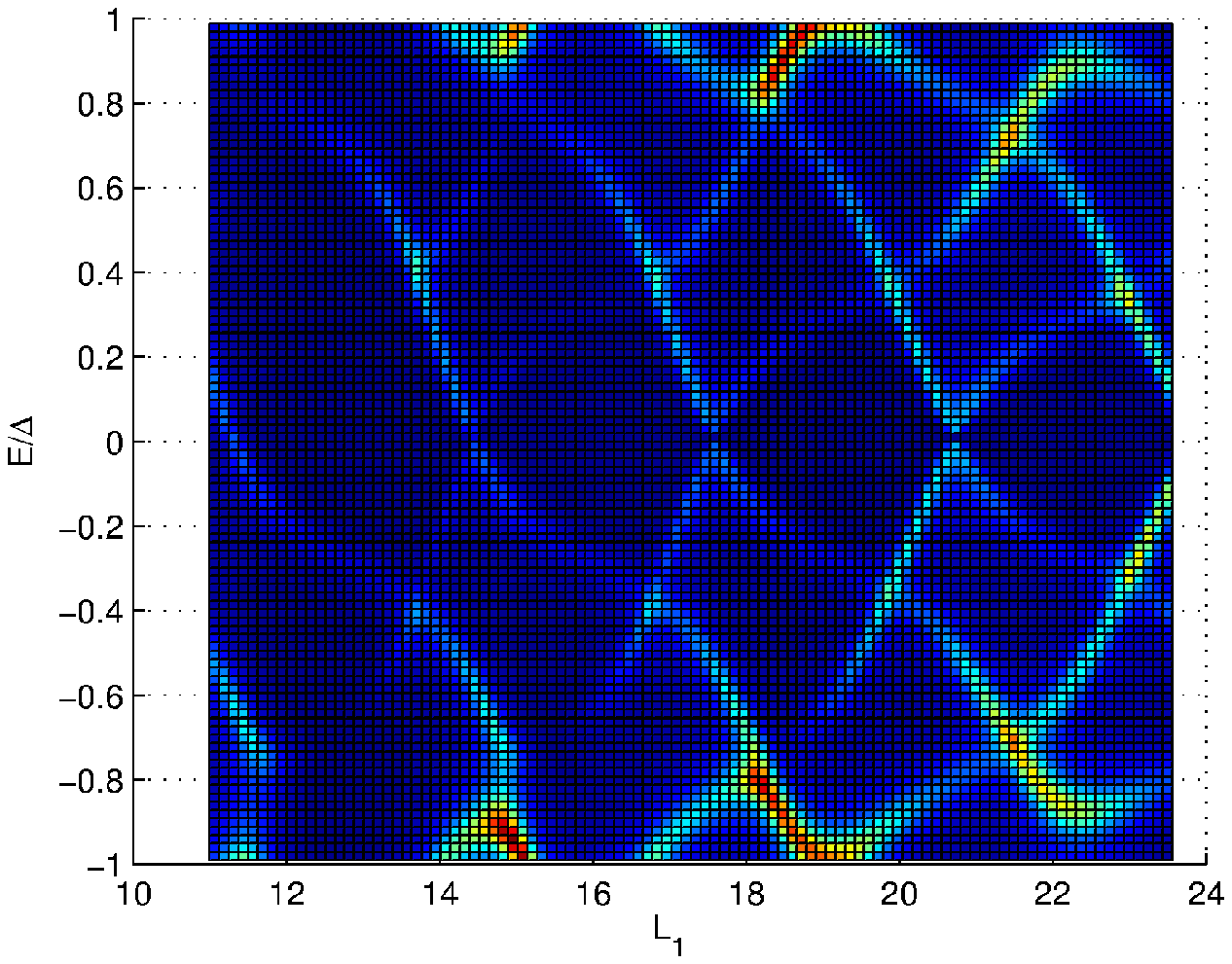}}
\subfigure[]{\ig[width=2.0in]{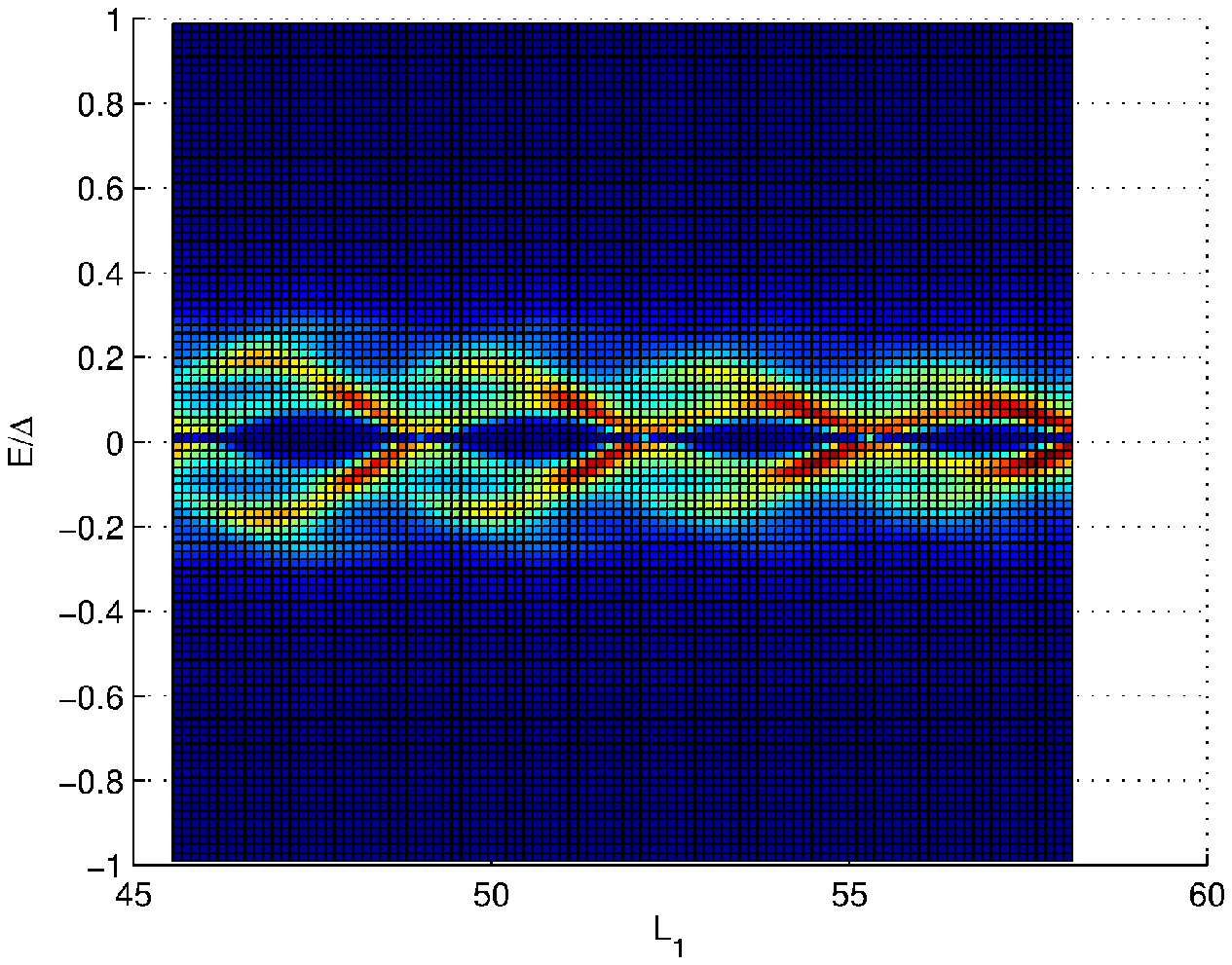}}
\subfigure[]{\ig[width=2.0in]{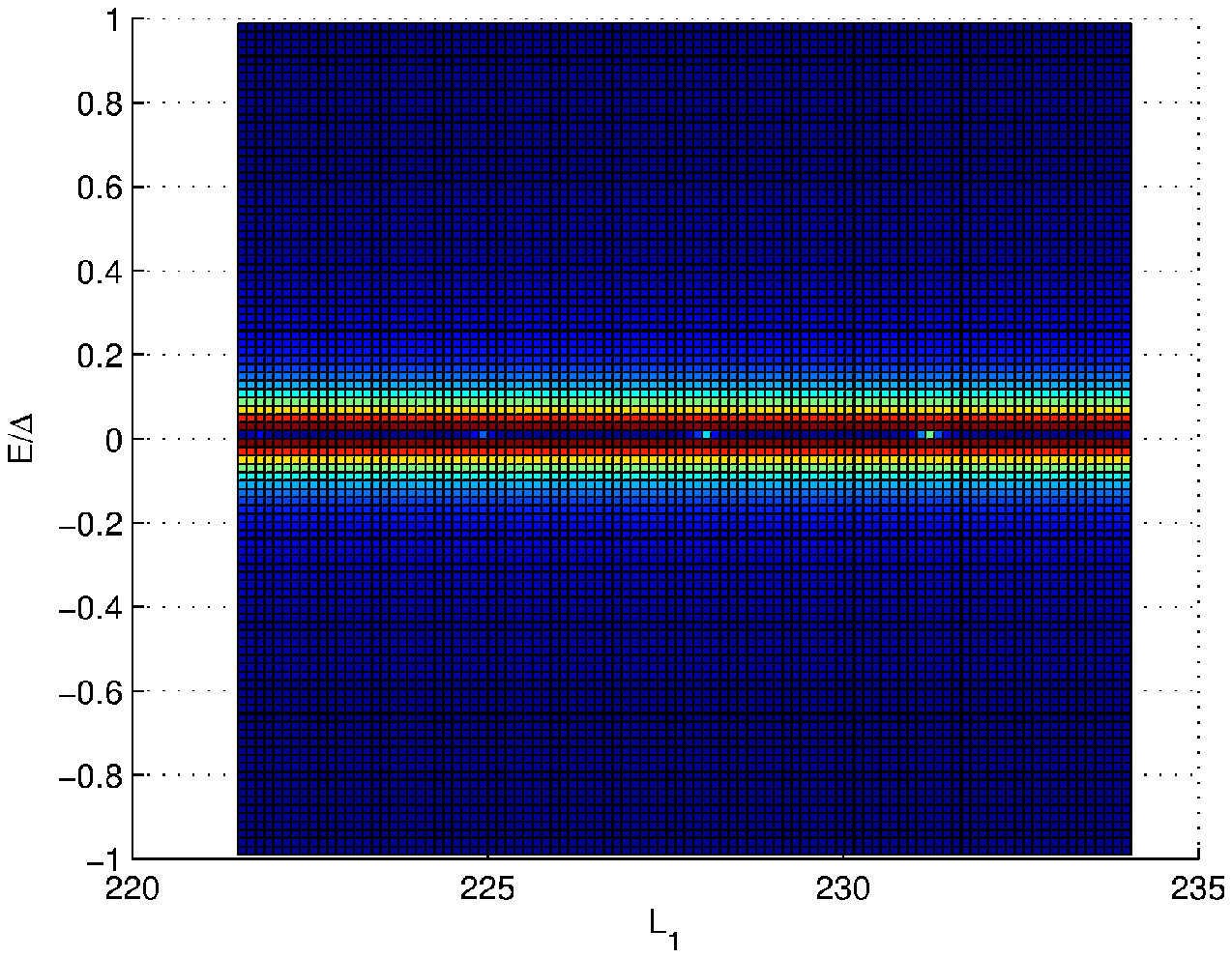}} \\
\subfigure[]{\ig[width=2.0in]{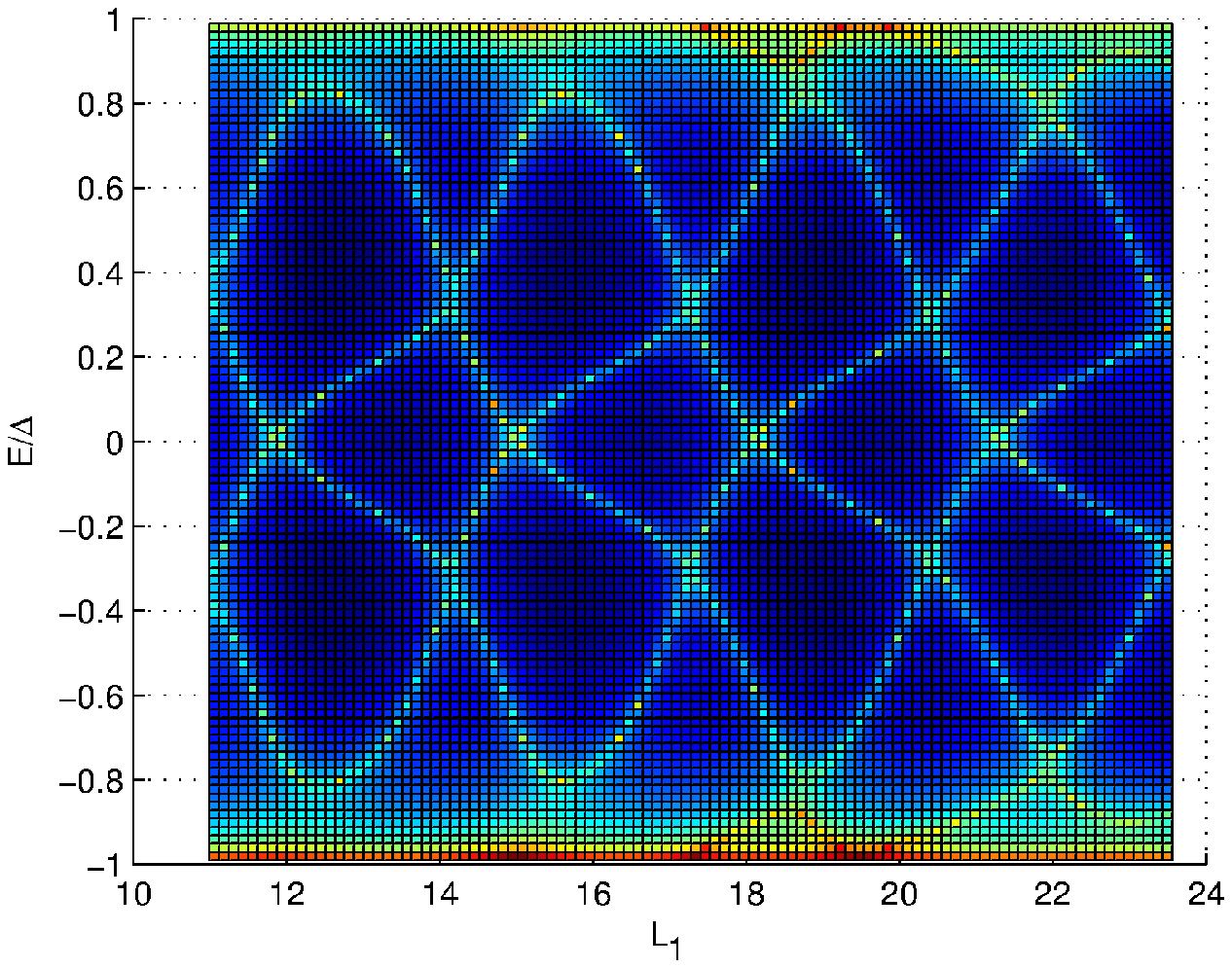}}
\subfigure[]{\ig[width=2.0in]{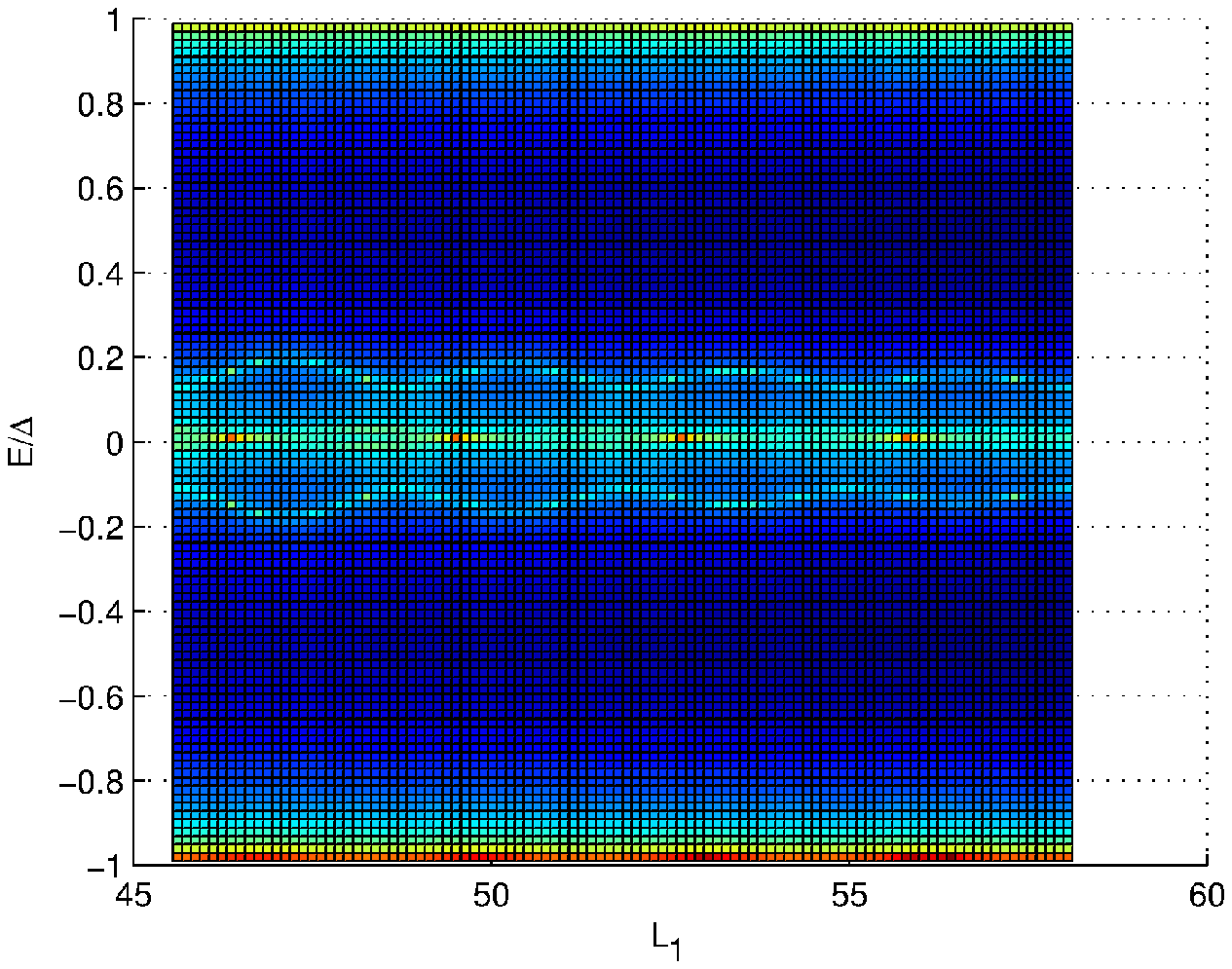}}
\subfigure[]{\ig[width=2.0in]{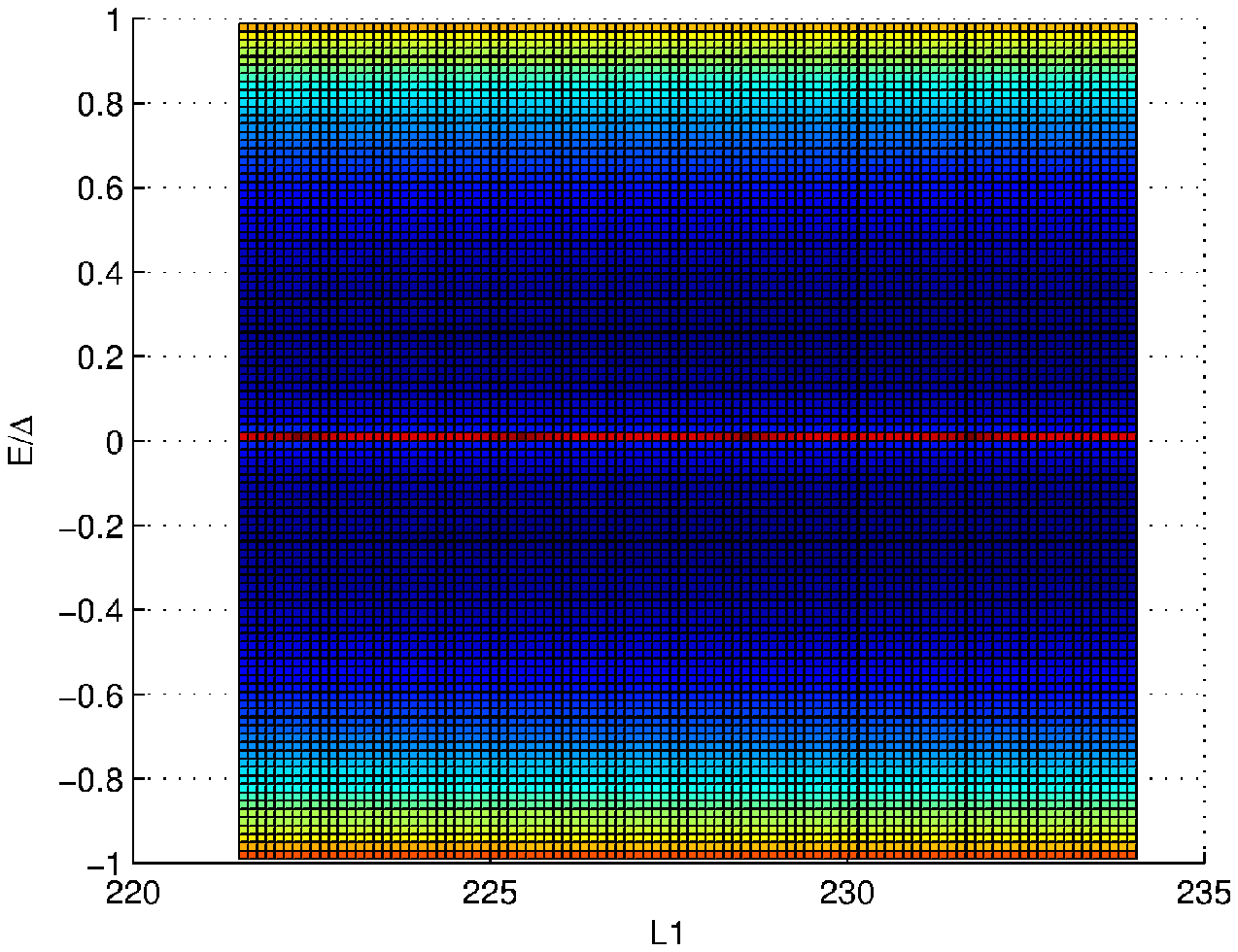}} \\
\subfigure[]{\ig[width=2.0in]{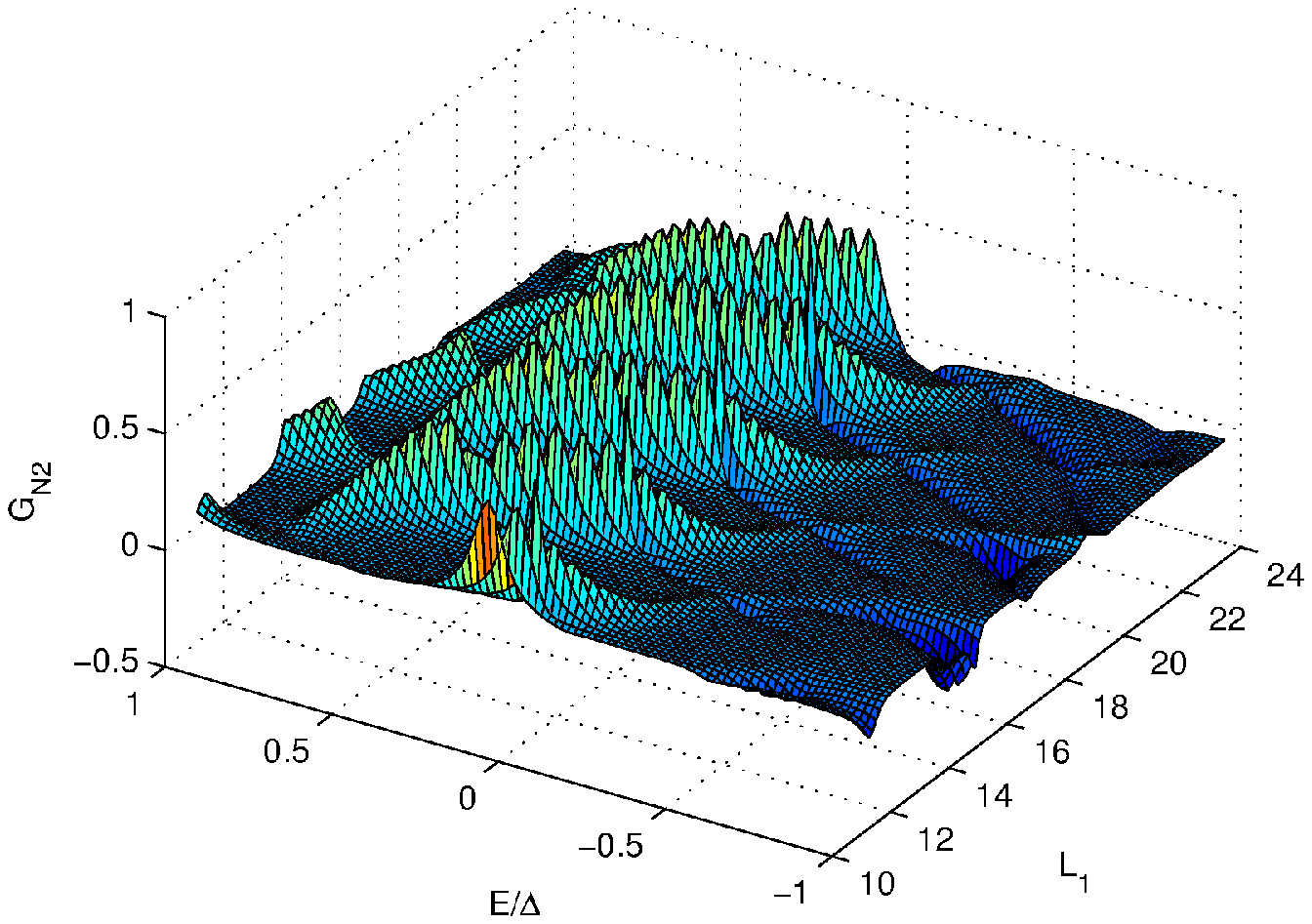}} 
\subfigure[]{\ig[width=2.0in]{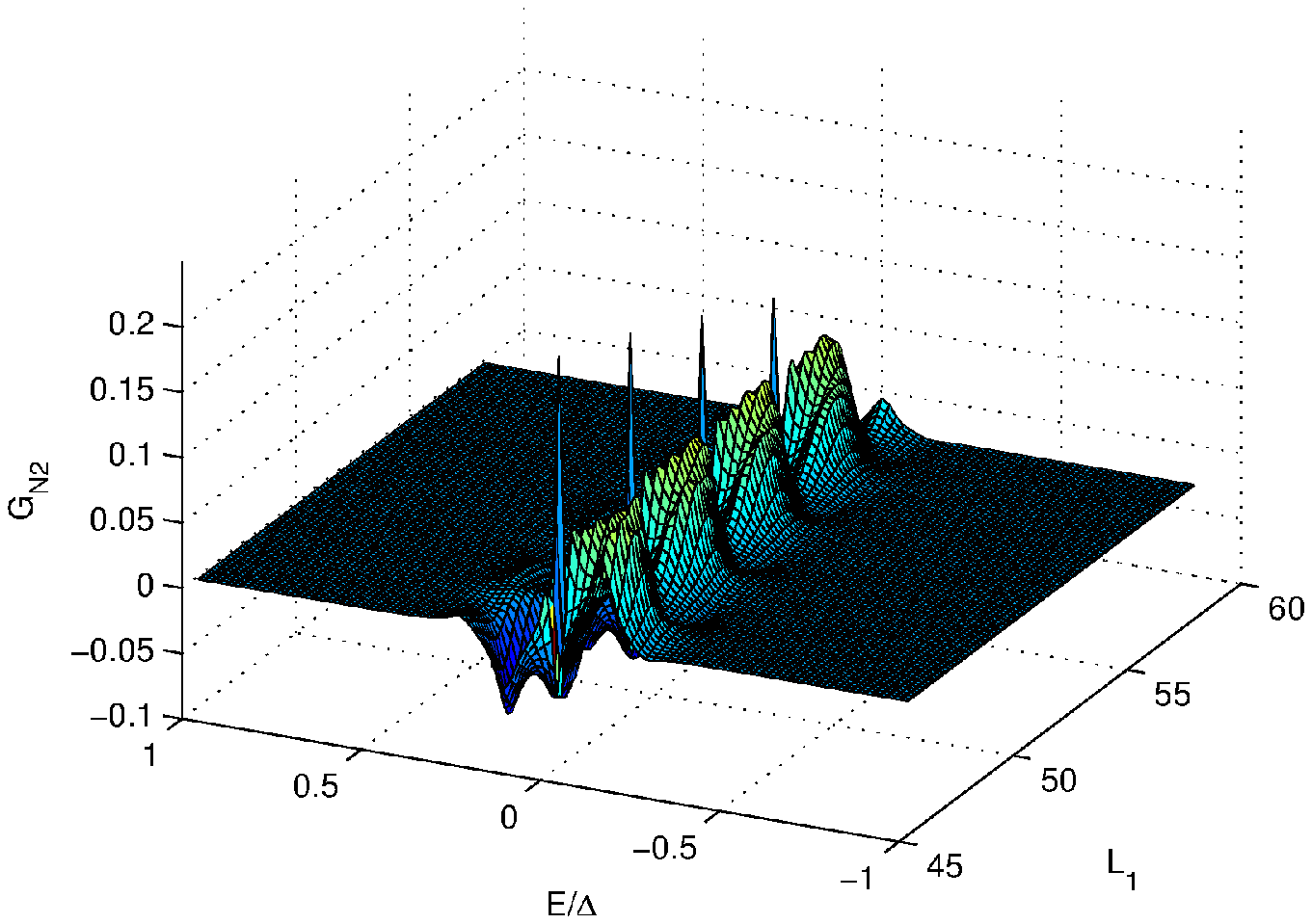}}
\subfigure[]{\ig[width=2.0in]{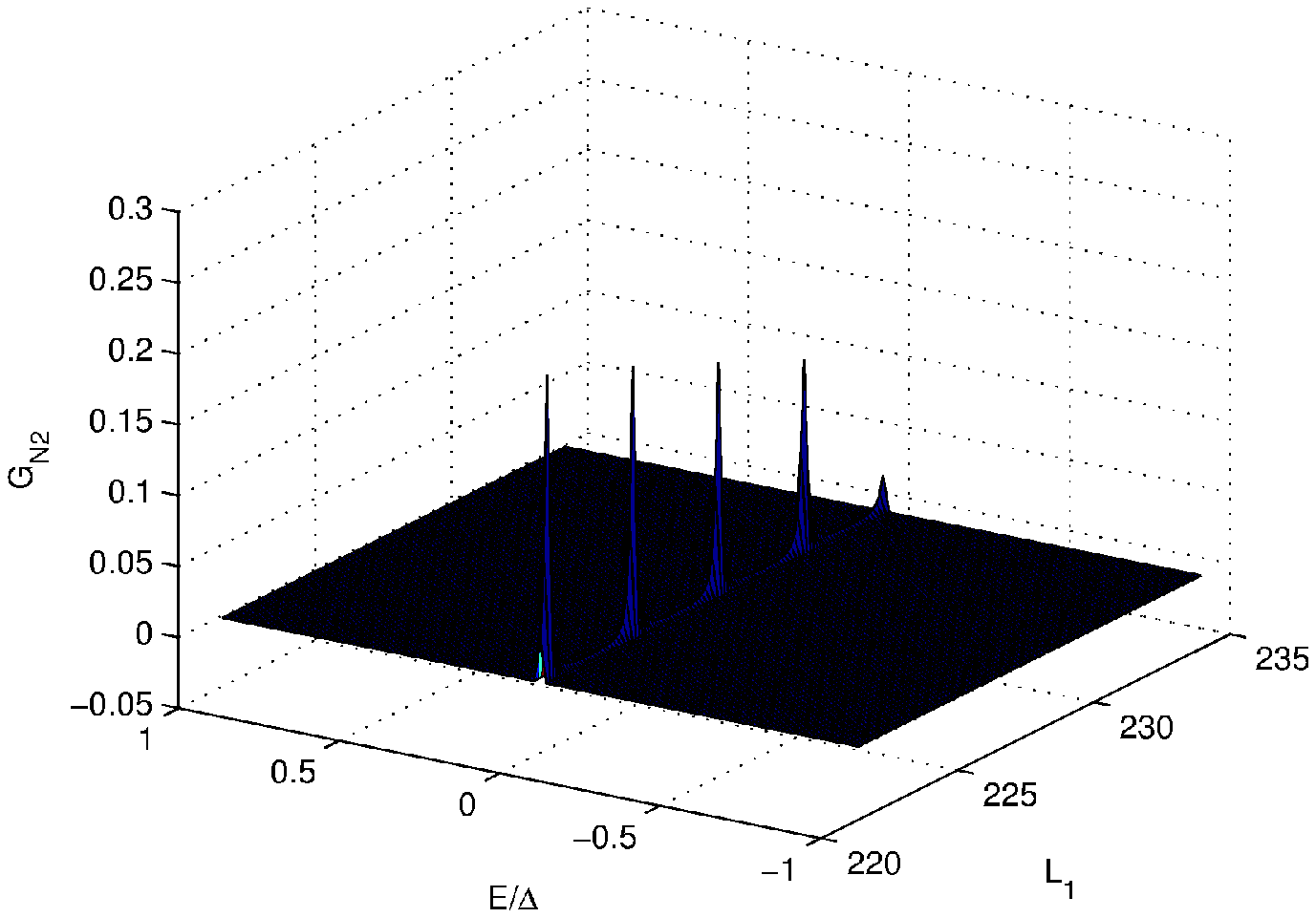}}
\caption{Plots of conductances and energies of sub-gap modes
when $\De_j$ has the same sign in the three SC wires: the parameters chosen 
are $k_F=1, ~m =0.5$ and $\De_1 = \De_2 = \De_3 = \De =0.1$, so that the 
length scale $\eta = 20$. In figures (a-c) and (g-i), the strength of the 
NM-SC barriers is $\la=5$, while in figures (d-f), we have $\la = \infty$
so that the NM leads do not play any role. In the first column, the length 
$L_1$ of the first SC wire varies from $3.5\pi$ to $7.5\pi$, while 
$L_2 =6.7\pi$ and $L_3 =6.3 \pi$ are held fixed; $L_j \simeq \eta$ in all 
cases. In the second column, $L_1$ varies from $14.5\pi$ to $18.5\pi$, while
$L_2 =12.3\pi$ and $L_3 =12.7\pi$; hence $L_j \simeq 2 \eta$.
In the third column, $L_1$ varies from $70.5\pi$ to $74.5\pi$, while
$L_2 =68.3\pi$ and $L_3 =68.7\pi$; hence $L_j \gg \eta$. The first 
row shows top views of surface plots of $G_C$ as a 
function of $L_1$ and $E/\De$ lying in the gap $-\De < E < \De$. 
The second row shows the energies $E/\De$ of the sub-gap modes as a function 
of $L_1$ for a system with just three SC wires; the energies are found by the 
vanishing of the determinant of a matrix as explained in the text. The peaks 
in $G_C$ shown in figures (a-c) match well with the sub-gap energies shown 
in figures (d-f). The third row shows surface plots of $G_{N2}$ as a function 
of $L_1$ and $E/\De$.} \label{fig:3sc2} \end{figure} \end{center} 
\end{widetext}

In the regime of large wire lengths, the normal conductance $G_{N2}$ shows 
some interesting properties. From the discussion above, we would conclude that 
$G_{N2}$ should be very small at this length regime. But in Fig.~\ref{fig:3sc2}
(i) we can see that $G_{N2}$ has discrete 
conductance peaks exactly at $E=0$ at some particular values of $L_1$,
and $G_{N2}$ is quite high at those points. If we increase $L_1$ more, the 
value of $G_{N2}$ decreases but the peaks still remain. This shows that the 
peaks are robust and are solely due to the three-wire geometry of the system 
as these kinds of peaks do not appear in a simple NM-SC-NM 
system~\cite{thakurathi}. We find that at the locations of these conductance 
peaks, the amplitudes $t_{1j}$,
$t_{2j}$, $t_{3j}$ and $t_{4j}$ take such values that $t_{1j}e^{-L_j/\xi}$,
$t_{2j}e^{-L_j/\xi}$, $t_{3j}e^{L_j/\xi}$ and $t_{4j}e^{L_j/\xi}$ are 
significantly larger compared to their values when there is no conductance 
peak. Hence for these values of the amplitudes, the wave functions inside
the SCs of the sub-gap modes are not negligible near the NM-SC junctions. So 
at these special points, an incoming electron from NM1 can easily enter the SC 
and then get transmitted to the NMs on the other sides by coupling to these 
sub-gap modes.

In the second column in Fig.~\ref{fig:3sc2}, we have taken $L_1=14.5\pi$ to 
$18.5\pi$, $L_2=12.3\pi$ and $L_3=12.7\pi$. In this intermediate length 
regime, $L_1$ is of the order of the decay length $\xi$. Hence, the sub-gap 
modes will now hybridize with each other and their energy will split from 
zero. But the hybridization will be less than that in the $L_1<\eta$ regime. 
Hence the energy splitting will also be less which 
is evident from the top view of $G_C$ in Fig.~\ref{fig:3sc2} (b). In that plot
we cannot see the six sub-gap modes separately. But there are states at 
non-zero energy due to the splitting, and the energy gap oscillates with 
$L_1$. The highest value of $G_C$ is lower than that in the $L_1 \gg \eta$ 
regime as the incoming electron now has a finite probability of transmitting 
to the other NM leads. 

The peaks in $G_{N2}$, as discussed in the large length regime, are present 
in this intermediate case also. But now $G_{N2}$ is non-zero at almost all 
values of $L_1$ because of the appreciable hybridization among the sub-gap 
modes. 

Let us now compare our results with the energy plots we get for a SC box made 
of three SC wires. In these plots we use the same length regimes to compare 
it directly with our results. In Fig.~\ref{fig:3sc2} (d), namely, 
when $L_1<\eta$, we see that for a fixed $L_1$ there are four points at 
different energies where $D=0$. These four points correspond to the four 
sub-gap modes in Fig.~\ref{fig:3sc2} (a). We cannot see the other two modes 
because they are very close to $E/\De= \pm 1$ and are therefore beyond our 
resolution. This difference between the numerical results for the conductances 
and for the determinant occurs because in the conductance calculation the 
barriers at the NM-SC junctions have a finite strength while the determinant 
is calculated for a SC box with infinitely large barriers to the NM leads. 

As we move to the intermediate length regime, Fig.~\ref{fig:3sc2} (e) shows 
that the $D=0$ lines approach zero energy; this occurs because the 
coupling between the sub-gap modes inside the SC box and therefore the 
splitting decreases. This is exactly what we see in the top view of $G_C$ in 
Fig.~\ref{fig:3sc2} (b). 

In the large length regime all the $D=0$ lines stay almost exactly at 
$E=0$ for all values of $L_1$ as we can see from Fig.~\ref{fig:3sc2} (f). Now 
the sub-gap modes in the SC box are almost decoupled. This matches with the 
top view of $G_C$ for large length, i.e., Fig.~\ref{fig:3sc2} (c).
We conclude that our numerical results for $G_C$ match well with the 
results we get for a SC box with infinite barriers at the three ends.

\vspace*{.6cm}
\noi {\bf 3.2 ~One of the $\De_j$'s with a different sign}
\vspace*{.4cm}

In this section we consider the case $\De_1<0$ and $\De_2, ~\De_3>0$, but we 
take the magnitudes of all the $\De_j$'s to be the same. As the electron is 
coming from the NM1 side, the results for $\De_1<0$ are different from 
the cases with $\De_2<0$ or $\De_3<0$. We will again calculate the 
conductances for three different length regimes and the determinant for a 
SC box. We then analyze the numerical results to see the differences in the 
conductances compared to the case when all the $\De_j$'s have the same
sign. We will take the same values of the parameters as before, so that 
the length scale $\eta=20$.

In the first column of Fig.~\ref{fig:3sc3}, we take $L_1=2.5\pi$ to $6.5\pi$,
$L_2=5.7\pi$ and $L_3=5.3\pi$. In this regime, from the top view of $G_C$ in 
Fig.~\ref{fig:3sc3} (a), we can clearly see that there are four sub-gap 
modes in the system. The sub-gap modes have non-zero energies as 
the decay length $\xi$ is greater than the length of the SCs. Instead of three
sub-gap modes at the junction of three wires, there is now only one sub-gap 
mode. In this length regime both $G_C$ and $G_{N2}$ are finite as we can see 
from the plots. We have already discussed the reason behind this earlier when 
all the $\De_j$'s have the same sign. 

In the third column, we choose $L_1=70.5\pi$ to $74.5\pi$, $L_2=68.3\pi$ and 
$L_3=68.7\pi$; hence $L_1 \gg \eta$ and is therefore much larger than the 
decay length. Similar to the earlier case when all the
$\De_j$'s are positive, here we again get $G_C$ almost equal to 2 
(in units of $e^2/h$) and $G_{N2}$ close to zero as the sub-gap modes 
are almost decoupled. The energy splitting of the sub-gap modes is also 
close to zero. Instead of having different energies, now all the 
four sub-gap modes lie close to zero energy as we can see from 
Fig.~\ref{fig:3sc3} (c). Near the ends of the gap, i.e., $E/\De=\pm 1$,
$G_{N2}$ rises which is due to the single particle density of 
states which is non-zero at those energies.
 
Unlike the case of all $\De_j$'s positive, $G_{N2}$ now does not show any 
conductance peaks at $E=0$ at any particular values of $L_1$. We have looked
at the numerical values of $t_{1j}$, $t_{2j}$, $t_{3j}$ and $t_{4j}$ at 
$E=0$ and at the values of $L_1$ where $G_{N2}$ had peaks for the case of
all $\De_j$'s positive. We find that $t_{1j}e^{-L_j/\xi}$,
$t_{2j}e^{-L_j/\xi}$, $t_{3j}e^{L_j/\xi}$ and $t_{4j}e^{L_j/\xi}$ are 
quite large in this case also. However $t_{n3}$ and $t_{a3}$ are almost 
equal; hence $G_{N2} = |t_{n3}|^2 - |t_{a3}|^2$ is almost equal to zero.
This is an interesting difference between the cases of all $\De_j$'s having
the same sign versus one of them having a different sign.

For the intermediate regime we have chosen $L_1=14.5\pi$ to $18.5\pi$, 
$L_2=12.7\pi$ and $L_3=12.7\pi$, so that $L_1 \simeq \eta$. In 
Fig.~\ref{fig:3sc3} (b) we see the that the sub-gap modes lie very 
close to $E=0$. There is a small energy splitting;
this splitting is due to the finite coupling of the sub-gap 
modes as the decay length is of the order of $L_1$. We find that $G_{N2}$ 
is almost zero and very small compared to $G_C$.

Next we consider a SC box and plot the determinant $D$ of the matrix of 
amplitudes as a function of $E/\De$ and $L_1$ as we did earlier, taking 
$\De_1<0$, and $\De_2, ~\De_3 > 0$. From the second row of Fig.~\ref{fig:3sc3}, 
we can see that the $D=0$ lines match well with our numerical results for 
the conductance peaks in the different regimes of length.

We have checked that instead of taking $\De_1<0$, if we choose $\De_2<0$ or 
$\De_3<0$ all the numerical results are qualitatively similar. 

\begin{widetext}
\begin{center} \begin{figure}[H]
\subfigure[]{\ig[width=2.2in]{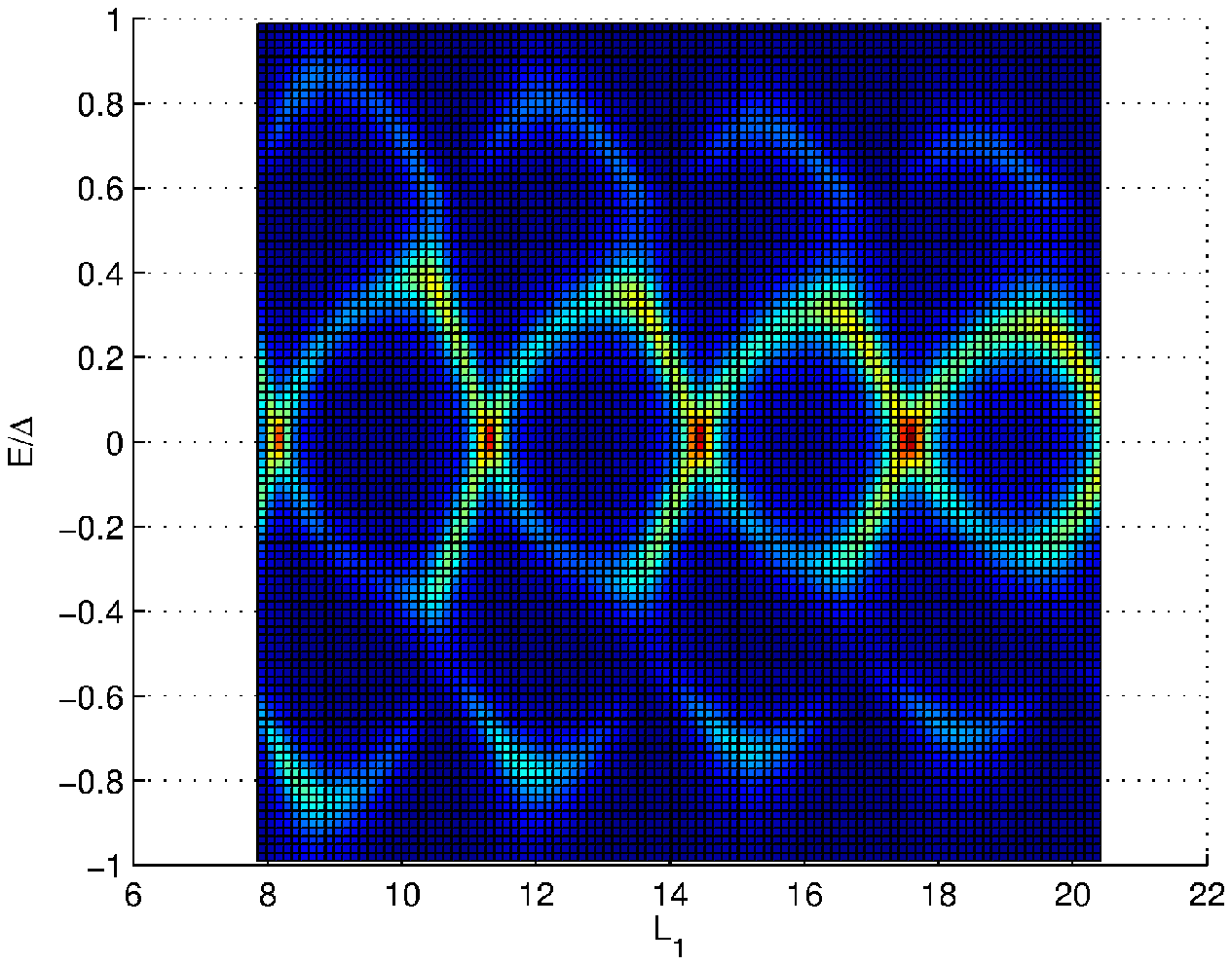}} 
\subfigure[]{\ig[width=2.2in]{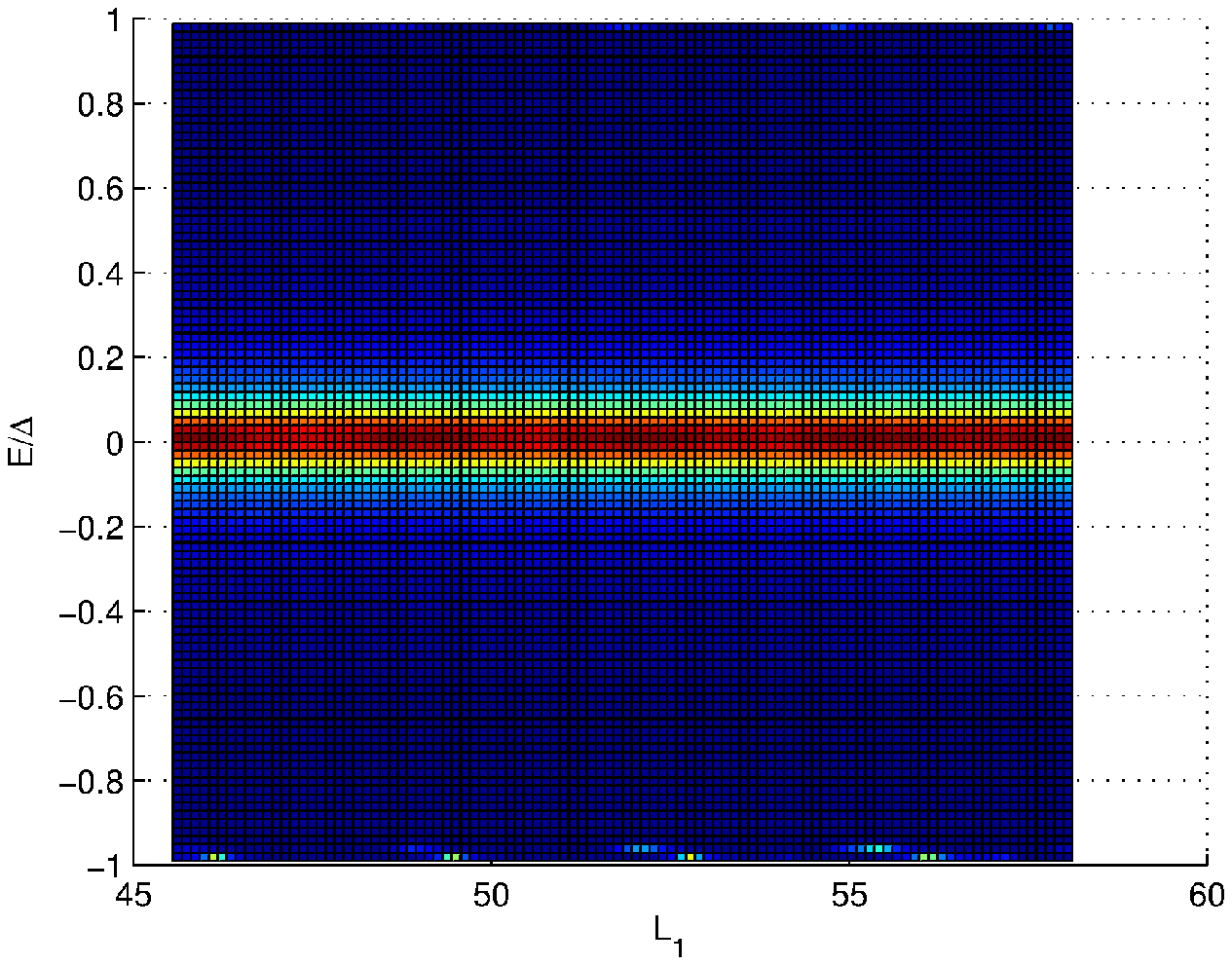}} 
\subfigure[]{\ig[width=2.2in]{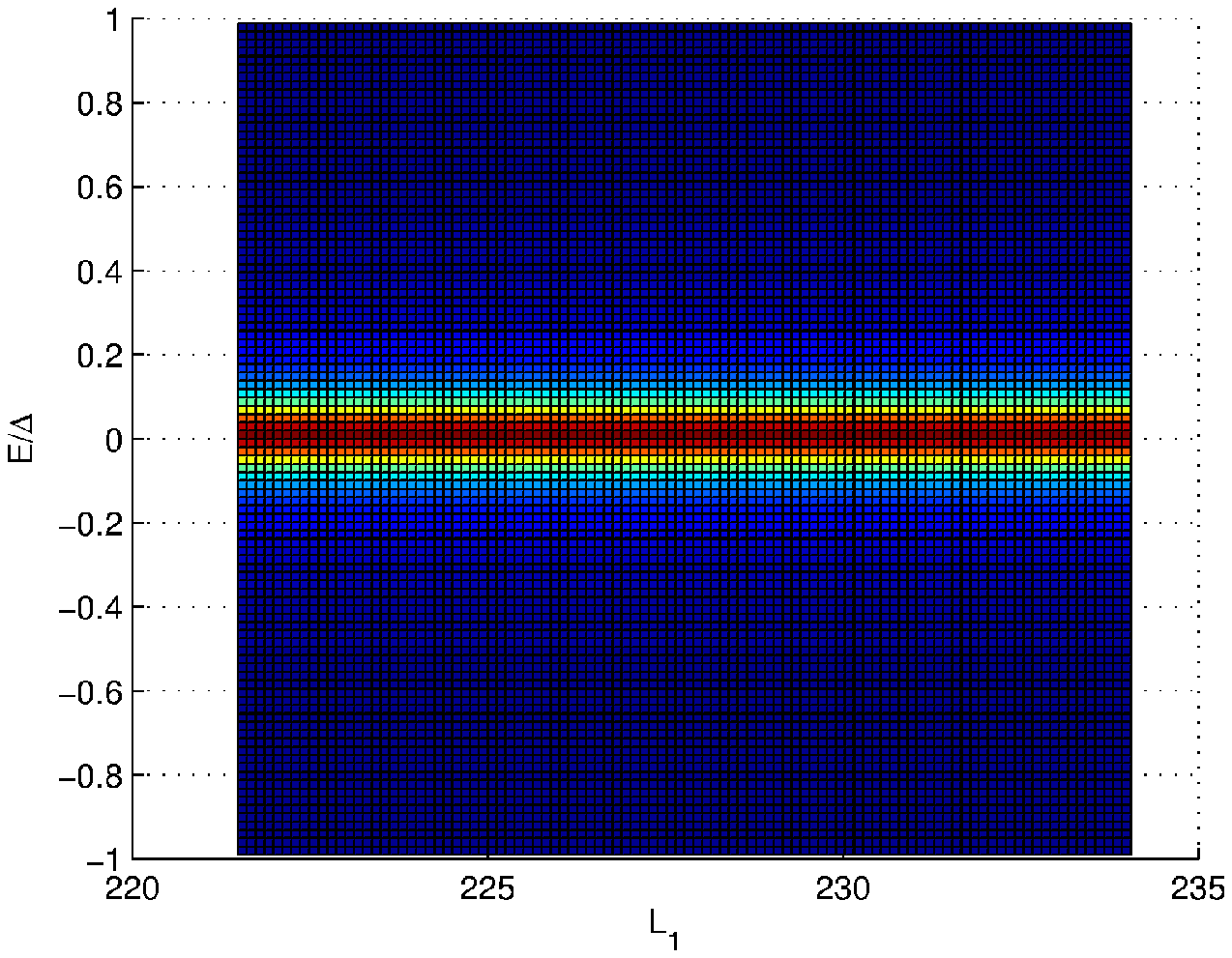}} \\
\subfigure[]{\ig[width=2.2in]{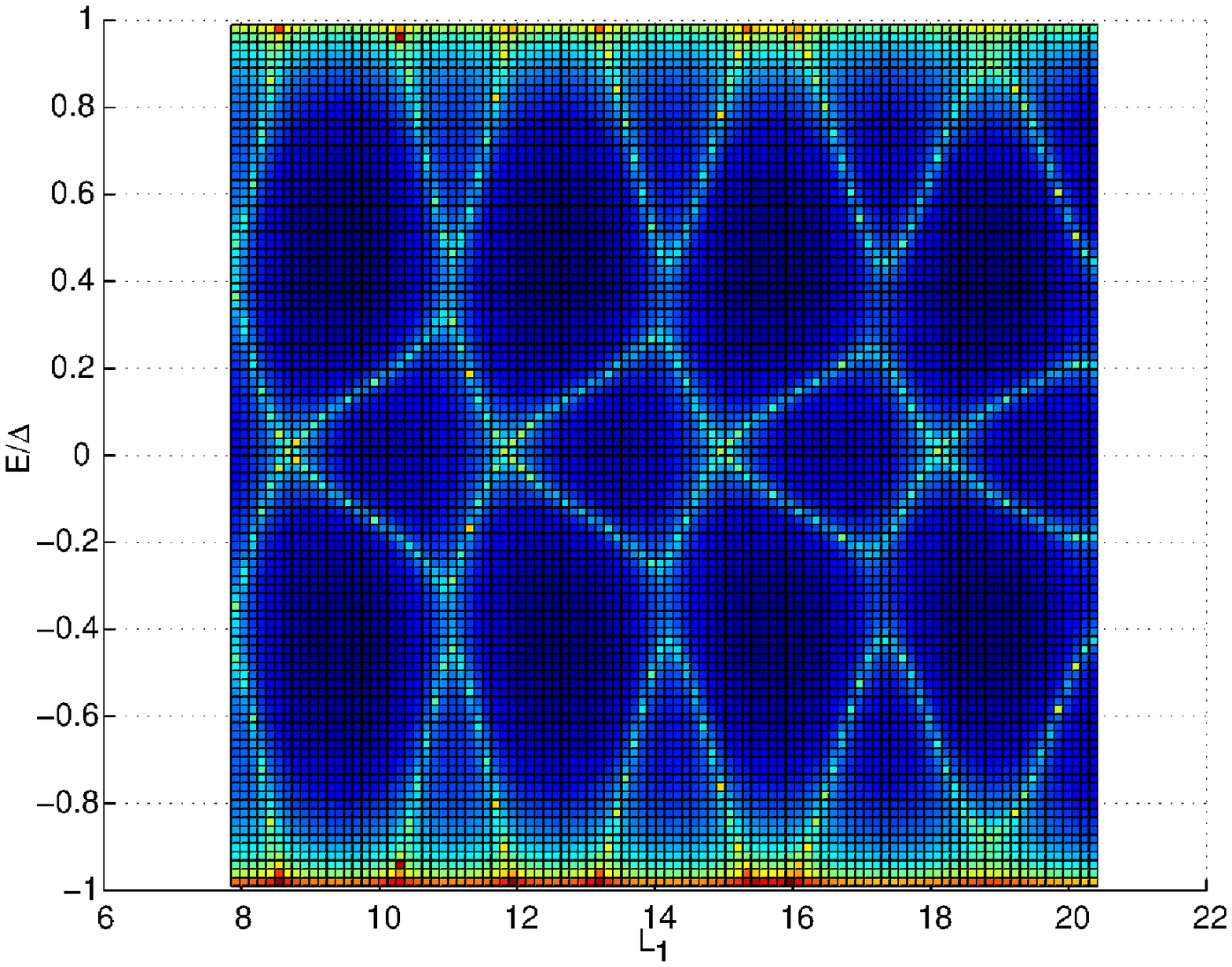}} 
\subfigure[]{\ig[width=2.2in]{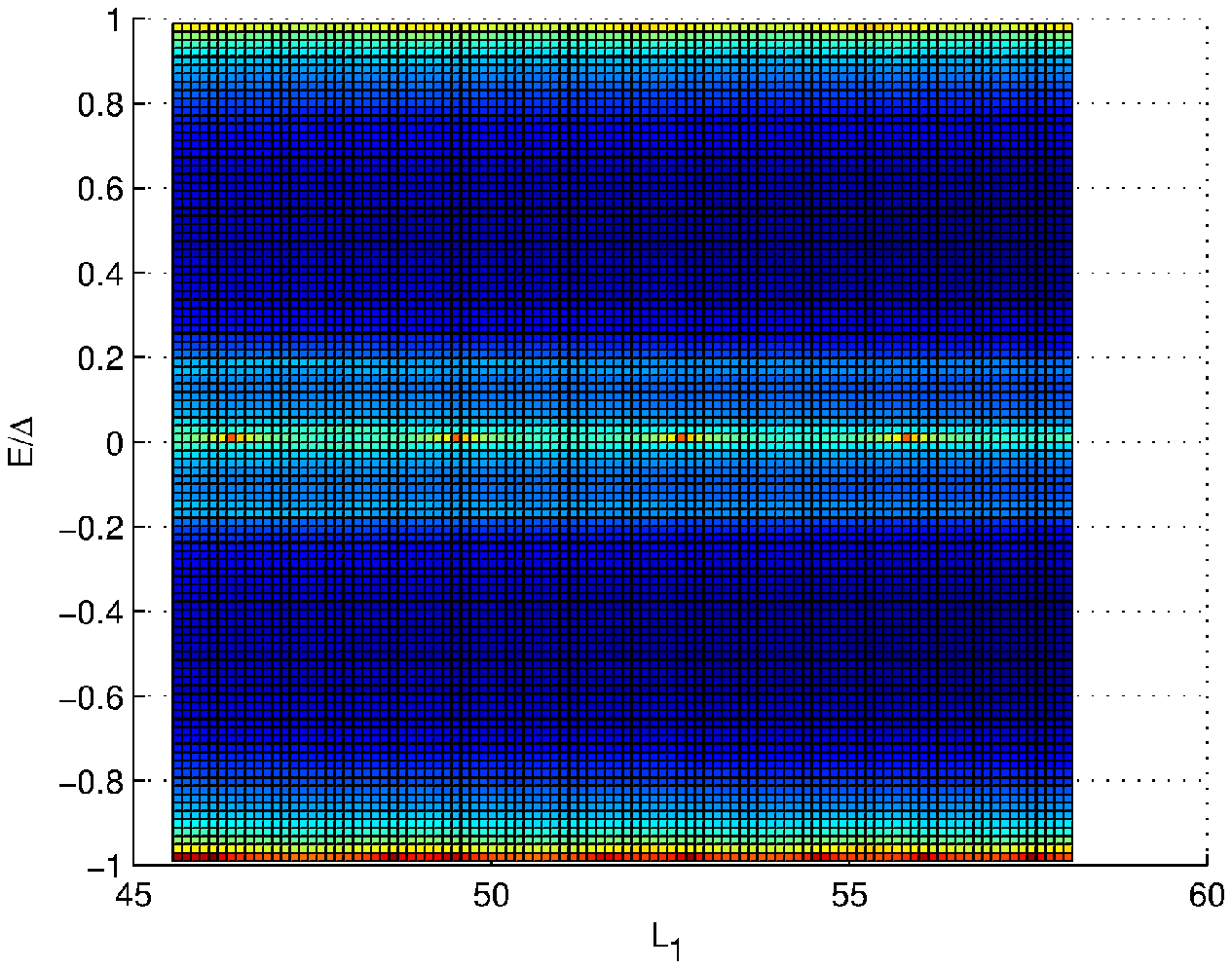}} 
\subfigure[]{\ig[width=2.2in]{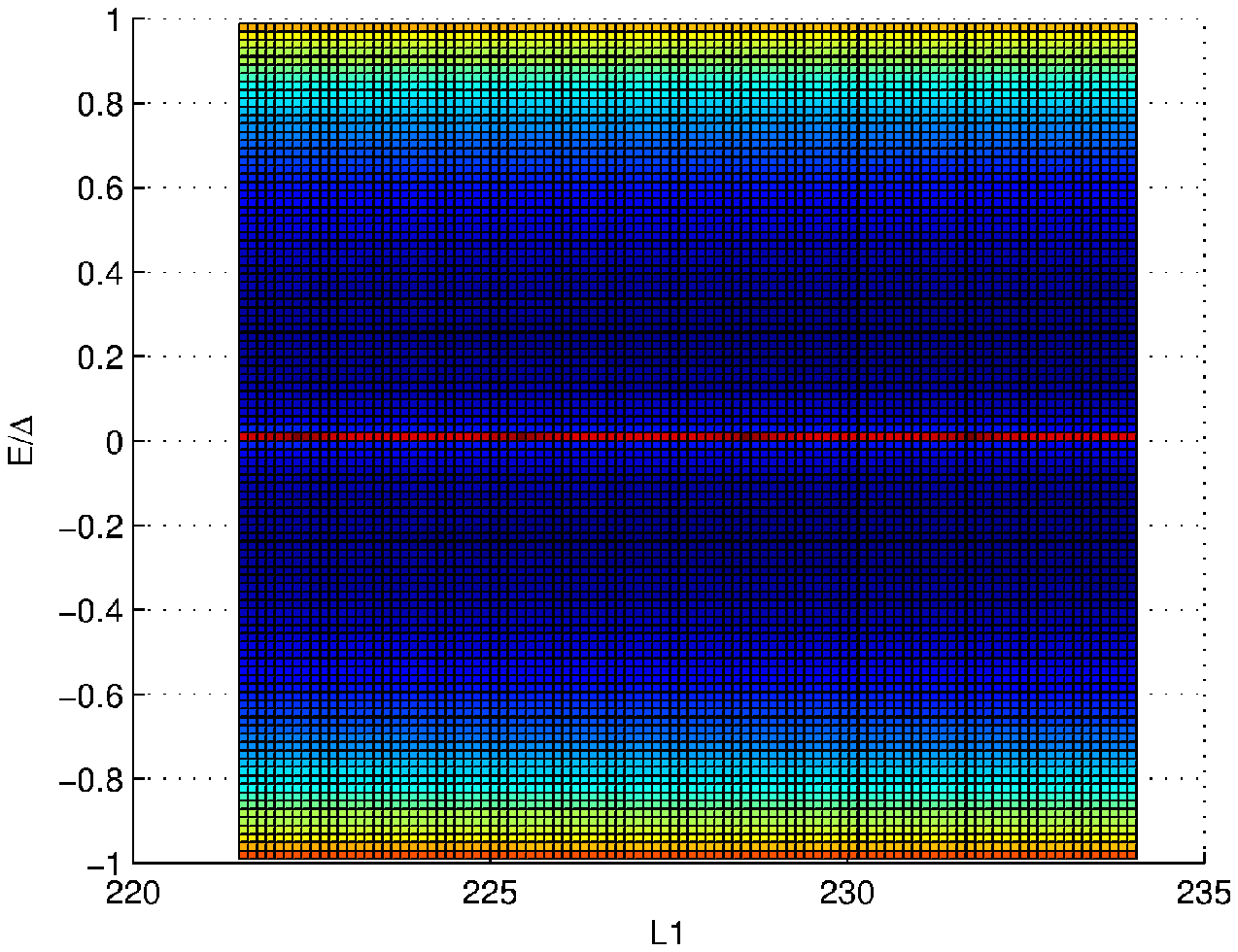}} \\
\subfigure[]{\ig[width=2.2in]{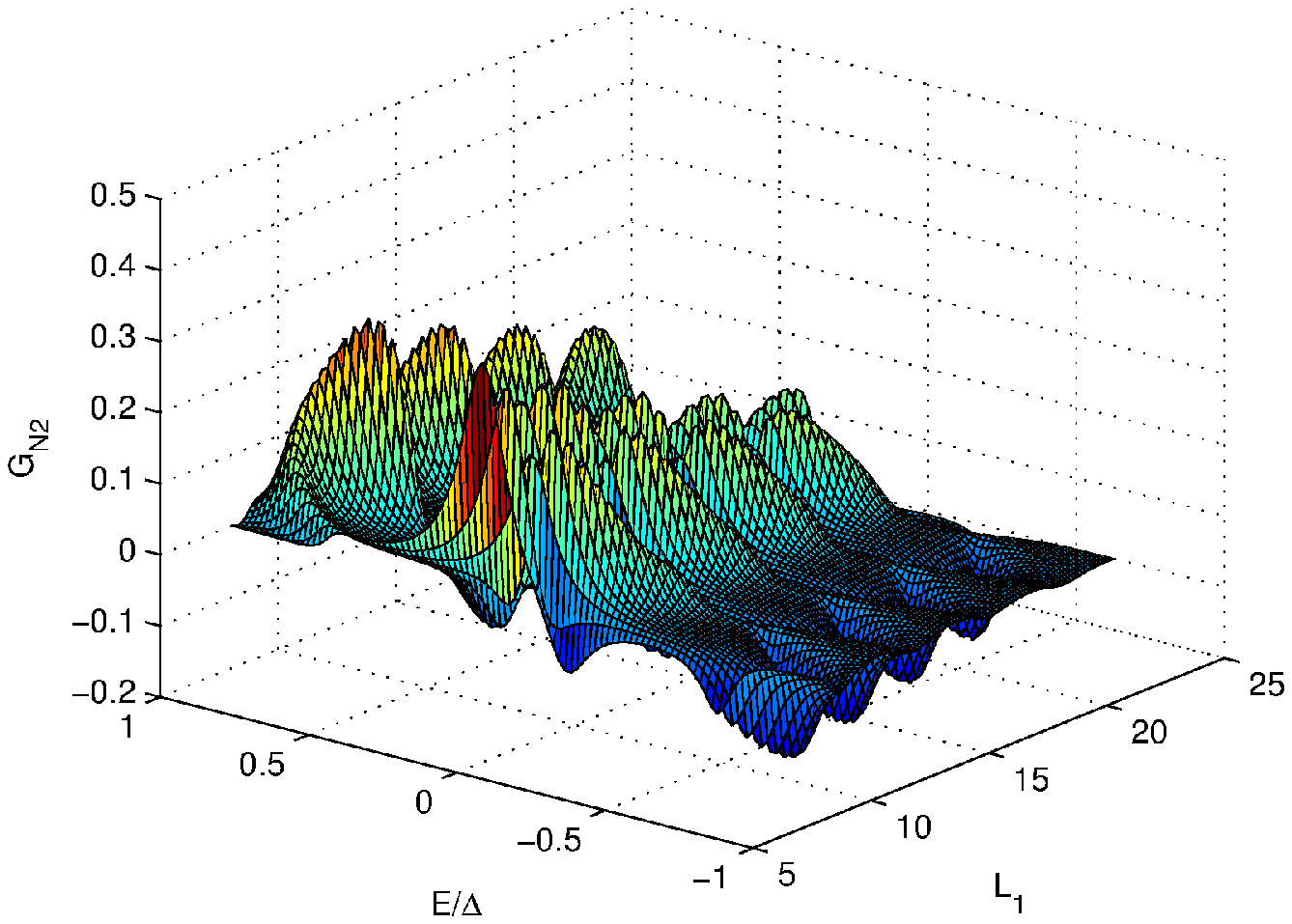}} 
\subfigure[]{\ig[width=2.2in]{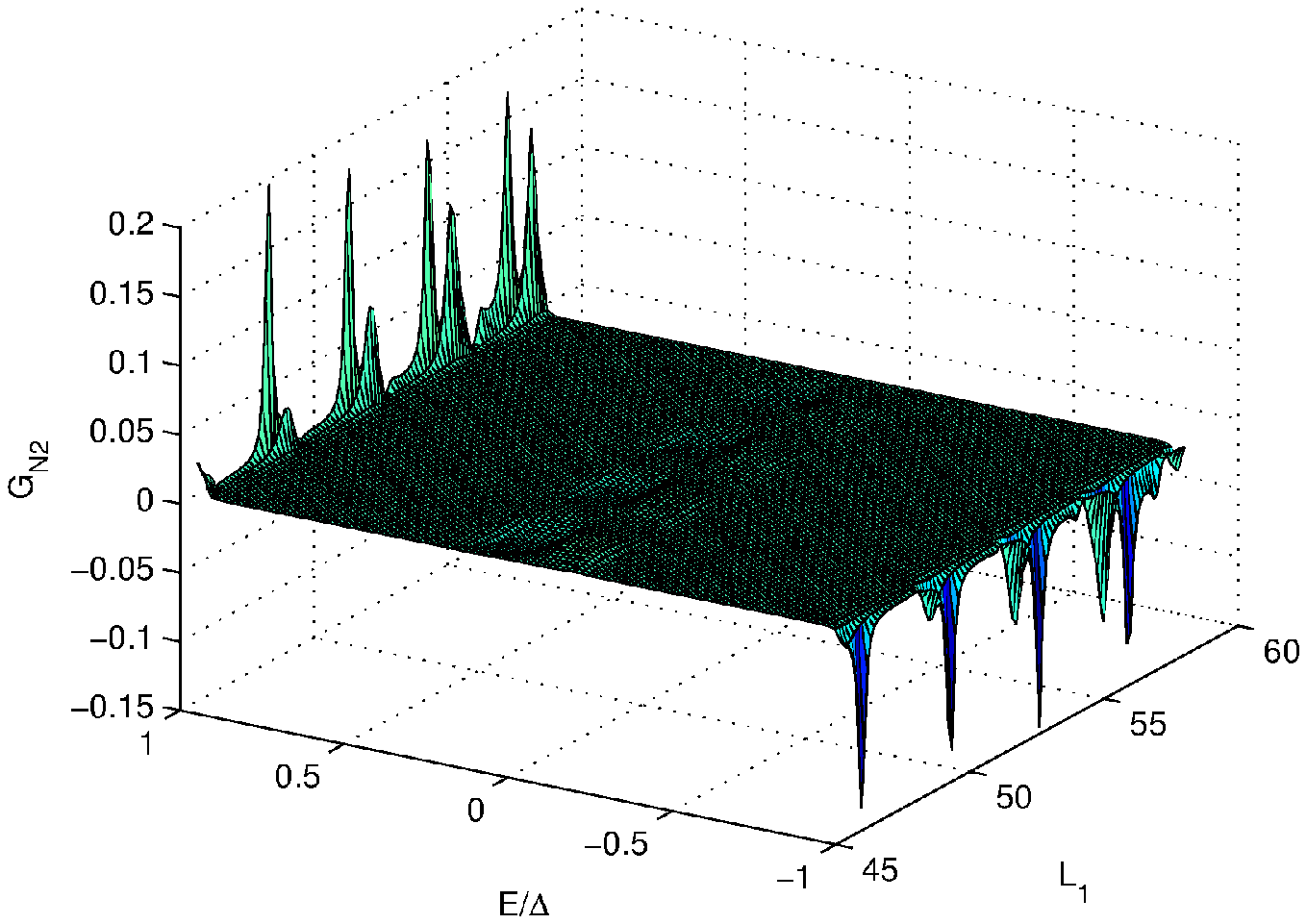}} 
\subfigure[]{\ig[width=2.2in]{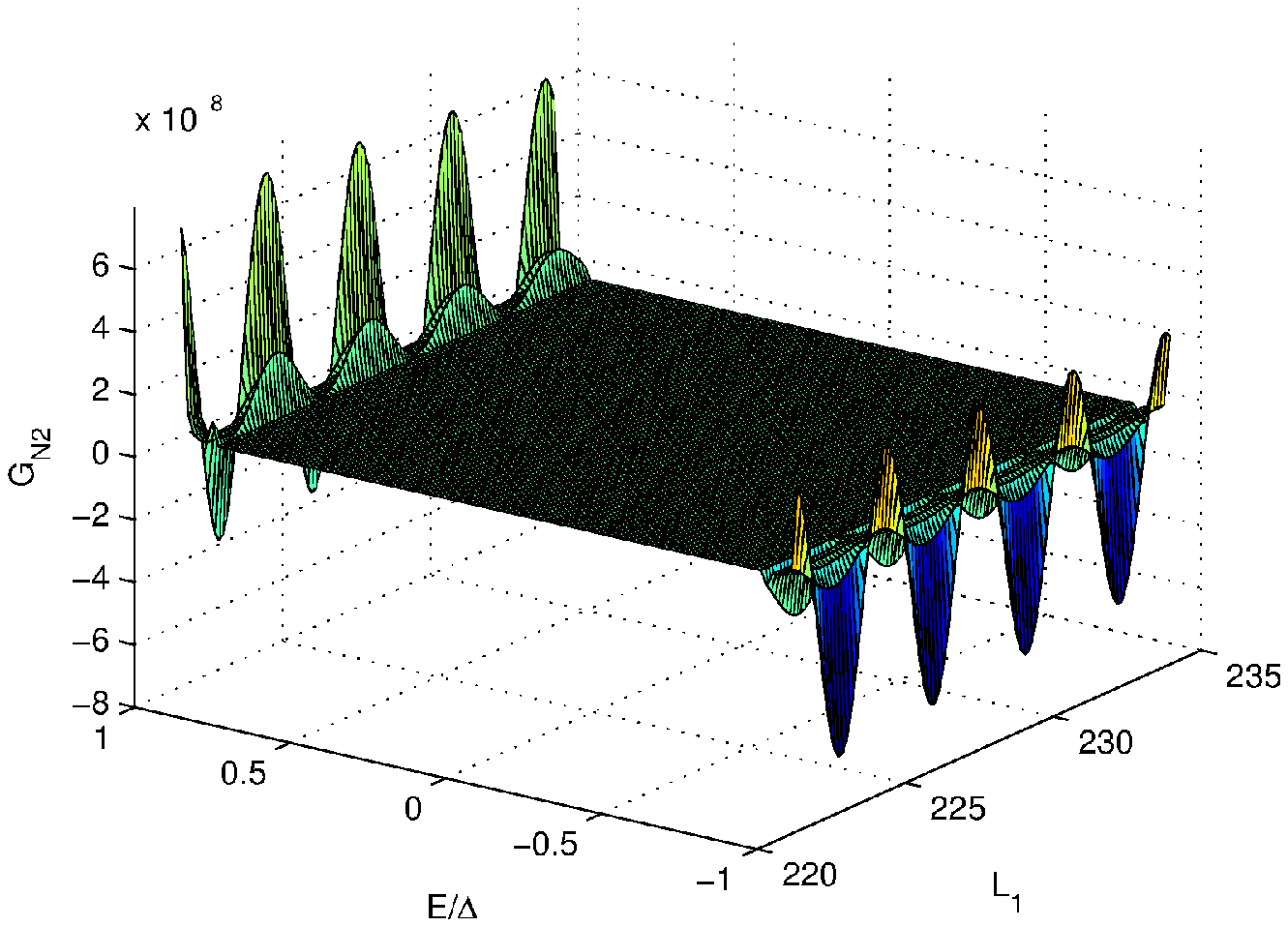}} 
\caption{Plots of conductances and energies of the sub-gap 
modes when $\De_1 < 0$, while $\De_2, ~\De_3 > 0$: the parameters chosen are 
$k_F=1, ~m =0.5$ and $-\De_1 = \De_2 = \De_3 = \De=0.1$, so that 
$\eta = 20$. In figures (a-c) and (g-i), the strength of the 
NM-SC barriers is $\la=5$, while in figures (d-f), we have $\la = \infty$
so that the NM leads do not play any role. In the first column, the length 
$L_1$ of the first SC wire varies from $3.5\pi$ to $7.5\pi$, while 
$L_2 =6.7\pi$ and $L_3 =6.3 \pi$ are held fixed; $L_j \simeq \eta$ in all 
cases. In the second column, $L_1$ varies from $14.5\pi$ to $18.5\pi$, while
$L_2 =12.3\pi$ and $L_3 =12.7\pi$; hence $L_j \simeq 2 \eta$.
In the third column, $L_1$ varies from $70.5\pi$ to $74.5\pi$, while
$L_2 =68.3\pi$ and $L_3 =68.7\pi$; hence $L_j \gg \eta$. The first 
row shows top views of surface plots of $G_C$ as a 
function of $L_1$ and $E/\De$ lying in the gap $-\De < E < \De$. 
The second row shows the energies $E/\De$ of the sub-gap modes as a function of
$L_1$ for a system with just three SC wires; the energies are found by the 
vanishing of the determinant of a matrix as explained in the text. The peaks 
in $G_C$ shown in figures (a-c) match well with the sub-gap energies
shown in figures (d-f). The third row shows surface plots of $G_{N2}$ as a 
function of $L_1$ and $E/\De$.} \label{fig:3sc3} \end{figure} \end{center}
\end{widetext}

We would like to point out that our results differ in many significant
ways from those in earlier papers (such as Ref.~\onlinecite{thakurathi})
which considered a single SC wire lying between two NM leads. In a single
SC wire, there are only two sub-gap modes lying at the ends of the wire;
they give rise to two peaks in the conductance.
If the $p$-wave pairing $\De$ changes sign somewhere inside the wire, two 
more sub-gap modes appear there; however these do not have an appreciable
effect on the conductances. In contrast to this, our system has several
sub-gap modes, three at the NM-SC junctions and one or three at the junction
of three SC wires depending on the relative signs of the $\De_j$'s. {\it All}
these sub-gap modes give rise to peaks in the conductance. To the best of our 
knowledge, this is the first model to be studied in which there are such
a large number of conductance peaks whose locations precisely match the
energies of all the sub-gap modes.

\vspace*{.6cm}
\noi {\large {\bf 4 ~Analytical understanding of Majorana modes at a 
junction of three SC wires}}
\vspace*{.4cm}

In this section we will study how many zero energy Majorana modes appear near 
a junction of three SC wires if the SC wires are semi-infinite and the NM 
leads are absent. Namely, the system only consists of three semi-infinite 
SC wires which meet at $x_j=0$.
Therefore out of the four scattering amplitudes $t_{1j}$, $t_{2j}$,$t_{3j}$ 
and $t_{4j}$, only those two will be non-zero for which the SC wave function 
$\psi_{scj} \to 0$ as $x_j \to \infty$. From Eqs.~\eqref{decay} 
and \eqref{grow}, we know that $\psi_{scj}$ is normalizable when only $t_{1j}$ 
and $t_{2j}$ are non-zero. Keeping this in mind we can write down the wave
functions for the semi-infinite wires.
\bea \psi_{scj}~ &=&~ t_{1j} ~e^{ik_1 x_j} ~\left( \begin{array}{c}
1 \\
sgn(\De_j) ~e^{i\phi} \end{array} \right) \non \\
&& + ~t_{2j} ~e^{-ik_2 x_j} ~\left(\begin{array}{c}
1 \\
-sgn(\De_j) ~e^{-i\phi} \end{array} \right). \label{semisc} \eea

We want to find Majorana modes which are at zero energy and are localized 
near the junction. For $E=0$, $e^{i\phi}= (E -i\sqrt{\De^2 -E^2})/\De$
reduces to $e^{i\phi}=-i$. Substituting this in Eq.~\eqref{semisc}, we get 
\bea \psi_{scj}~ &=&~ t_{1j}e^{ik_1 x_j}\left( \begin{array}{c} 
1 \\
-i ~sgn(\De_j) \end{array} \right) \non \\
&& + t_{2j} e^{-ik_2 x_j} \left(\begin{array}{c}
1 \\
-i ~sgn(\De_j) \end{array} \right). \label{semisc1} \eea 
Now we use the boundary conditions in Eqs.~\eqref{bc9} to find the Majorana 
modes. Using \eqref{bc9}, for the $c_j$'s we get
\bea && ik_1~t_{1j} ~-~ ik_2~t_{2j} \non \\
&& = ~M_{j1} ~(t_{11}+t_{21}) ~+~ M_{j2} ~(t_{12}+t_{22}) \non \\ 
&& ~~+~ M_{j3} ~(t_{13}+t_{23}) \non \\
&& ~~-~ \frac{m |\De_j |}{k_F}(t_{1j}+t_{2j}), \label{ex} \eea
while for the $d_j$'s we get, 
\bea && sgn(\De_j) ~[ik_1~t_{1j}-ik_2~t_{2j}] \non \\ 
&& =~ M_{j1} ~sgn(\De_1) ~(t_{11}+t_{21}) \non \\ 
&& ~~+~ M_{j2} ~sgn(\De_2) ~(t_{12}+t_{22}) \non \\ 
&& ~~+~ M_{j3} ~sgn(\De_3) ~(t_{13}+t_{23}) \non \\ 
&& ~~-~ \frac{m|\De_j| ~sgn(\De_j)}{k_F} ~(t_{1j}+t_{2j}). \label{hx} \eea

So we have a total of six equations, three for the $c_j$'s and three for 
the $d_j$'s. From Eqs.~\eqref{ex} and \eqref{hx} it can be shown that if all 
the $\De_j$'s have the same sign, the equations for the $c$'s and $d$'s are 
exactly the same. So instead of six equations we get only three independent 
equations. But we have six variables in this problem since we have three wires 
and each of the wires has two different scattering amplitudes corresponding to 
$k_1$ and $k_2$. Hence we have three equations for six variables. Hence, 
there are three independent solutions and therefore three Majorana modes 
at the junction.

Now we consider the case when one of the $\De_j$'s has a different sign
from the other two; let us assume that $\De_2$ and $\De_3$ have the same sign 
while $\De_1$ has the opposite sign. In this case, it is straightforward to 
show that $t_{11}+t_{21}=0$. 
So we now have five independent variables instead of six. Using the above 
relation it can be shown that the equations for $c_2$ and $c_3$ are the 
same as the equations for $d_2$ and $d_3$ respectively. But the equations
for $c_1$ and $d_1$ are different. So we have four independent equations and 
five variables which implies that we have only one independent solution
and hence only one Majorana mode. 

These are the only two combinations of the signs of $\De_j$'s which will give 
different results. Eq.~\eqref{dej} implies that any other combination will be 
similar to either the first case or the second case. To summarize, the two 
cases are as follows.

\noi (i) If all three $\De_j$'s have the same sign then there are three 
Majorana modes at the junction.

\noi (ii) If one of the $\De_j$'s has the opposite sign to the other two 
then there is only one Majorana mode at the junction. 

The existence of one or three zero energy Majorana modes can also be 
understood as follows. A well known lattice model of a $p$-wave 
superconducting wire is the Kitaev chain~\cite{kitaev1}. We first consider
a single Kitaev chain. The electron operators $c_n$ and $c_n^\dg$ at site $n$ 
can be written in terms of two Hermitian operators, called $a_n$ and $b_n$, as
$c_n = (1/2) (a_n + i b_n)$ and $c_n^\dg = (1/2) (a_n - i b_n)$. (These 
operators satisfy the anticommutation
relations $\{ a_m , a_n \} = \{ b_m , b_n \} = 2 \de_{mn}$). The 
Hamiltonian of the Kitaev chain only has terms of the form $ia_m b_n$, not 
$i a_m a_n$ or $i b_m b_n$. Such a Hamiltonian has an ``effective time reversal
symmetry", namely, it is invariant under complex conjugation of all complex 
numbers along with $a_n \to a_n$ and $b_n \to - b_n$~\cite{degottardi2}. 
This symmetry implies 
that if we look at eigenstates with zero energy, their wave functions involve 
only the $a_n$ or only the $b_n$, not both. A long Kitaev chain has zero energy
Majorana modes at the two ends. Depending on the sign of the $p$-wave pairing,
positive or negative, it turns out that the Majorana mode at the left end is
of type $a$ and at the right end is of type $b$ or vice 
versa~\cite{degottardi2}. We now consider three semi-infinite Kitaev chains 
which meet at one site; the three Kitaev chains could have the same or 
different signs of the $\De_j$'s. If all the $\De_j$'s have the same sign, 
the three Majorana modes at the junction will all be of the same type, say, 
$a$. Then they will not mix with each other since the Hamiltonian has no terms
of the form $i a_m a_n$; hence they will remain at zero energy. On the other 
hand, if one of $\De_j$'s has a different sign from the other two, two of the 
Majorana modes will be of one type, say $a$, and the other will be of type $b$.
Now the $b$ mode can mix with the two $a$'s; as a result the energies of 
two of the Majorana modes will become non-zero, namely $\pm E$, while one 
Majorana mode will stay at zero energy. (Written in terms of the $a$ and $b$ 
operators, the Hamiltonian of the system can be seen to have an $E \to - E$ 
symmetry; hence states can mix and move away from zero energy only in pairs).

Having seen that a Kitaev chain has two kinds of Majorana end modes,
called $a$ and $b$, we can write down an effective Hamiltonian which describes
the sub-gap physics of our three-wire system. This Hamiltonian will only 
contain the operators for the Majorana end modes (the operators in the rest 
of the SC wires are not required) and the electron operators in the NM 
wires~\cite{fu,fidkowski2,hutzen,affleck}. The Hamiltonian will have three
kinds of terms. First, the Hamiltonian at the junction of the 
three SC wires can be written as follows. If 
all three Majorana modes are of the same type, then no coupling between
them is allowed due to the effective time reversal symmetry. But if 
two of the modes are of one type (say, type $a$ on wires 1 and 2) and the 
third mode is of the other type ($b$ on wire 3), then this part of the
Hamiltonian will take the form $H_1 = i (w_1 a_1 + w_2 a_2) b_3$, where
$w_1$ and $w_2$ are two couplings (assumed to be real). Next,
on each wire $j$, the Majorana mode near the junction with the NM wire is
of the opposite type to the mode near the junction of the three SC wires;
namely, at the junctions with NM wires, the modes will be given by $b_1$,
$b_2$ and $a_3$. The hybridization between the two modes at the ends of
each wire will give rise to three more couplings given by $H_2 = i (w_3
a_1 b_1 + w_4 a_2 b_2 + w_5 a_3 b_3)$. (The couplings $w_3$, $w_4$ and
$w_5$ go to zero exponentially as the lengths of the SC wires become
large). Finally, we have couplings at 
the junctions of the SC and NM wires between the Majorana modes at $x_j = 
L_j - \ep$ and the electron operators in the NM wires at $x_j = L_j + 
\ep$ (following the notation in Eq.~\eqref{Ham}). We can think of this 
coupling as arising from a hopping between the operator $c_{sc} = (1/2) 
(a + ib)$ at $x_j = L_j - \ep$ in the SC wire and the operator $c_{nm}$ at 
$x_j = L_j + \ep$ in the NM wire, namely, a term of the form $c_{sc}^\dg c_{nm}
+ c_{nm}^\dg c_{sc}$. If the Majorana mode is of type $a$, we set $b=0$ and get
$c_{sc}^\dg c_{nm} + c_{nm}^\dg c_{sc} = (1/2) a (c_{nm} - c_{nm}^\dg)$, 
while if the Majorana is 
of type $b$, we set $a=0$ and get $c_{sc}^\dg c_{nm} + c_{nm}^\dg c_{sc} = 
-(i/2) b (c_{nm} + c_{nm}^\dg)$. We thus get a coupling of the form 
\bea H_3 &=& i w_6 ~b_1 ~[c_{nm} (L_1 + \ep) + c_{nm}^\dg (L_1 + \ep)] \non \\
&& + ~i w_7 ~b_2 [c_{nm} (L_2 + \ep) + c_{nm}^\dg (L_2 + \ep)] \non \\
&& +~ w_8 ~a_3 [c_{nm} (L_3 + \ep) - c_{nm}^\dg (L_3 + \ep)]. \eea
The complete effective Hamiltonian is given by the sum of $H_1$, $H_2$ and
$H_3$ and has eight parameters $w_i$.

The parameters $w_i$ can be determined as follows. We first use the 
microscopic 
Hamiltonian in Eq.~\eqref{Ham}, the $\de$-function potentials $\la$ at the 
NM-SC junctions and the matrix $\bf M$ at the junction of three SCs to 
compute the energies and wave functions of all the sub-gap modes. We then
fit the values of these quantities with those calculated from the 
effective Hamiltonian described above in order to find the values of $w_i$. 
We will not carry out this exercise here. The effective Hamiltonian can then 
be used to compute the various conductances of the system~\cite{fidkowski2}.

\vspace*{.6cm}
\noi {\large {\bf 5 ~Renormalization group equations for a junction of 
several wires}}
\vspace*{.4cm}

In this section, we will consider the effect of electron-electron interactions
on the conductances of a system of two or more NM wires which meet at a 
junction where there is a SC region (we can consider this region to be a SC 
dot). This system differs from the one studied in the earlier sections in two 
major ways. First, the region of three SC wires considered in Sects. 2 - 4 
will now be taken to be a single region. Further, the only role played by this 
region will be to give rise to a scattering matrix $S$ for the NM wires. Due 
to the superconductivity, we have to consider the cases of both incident 
electrons and incident holes, and both normal and Andreev reflection and 
transmission processes; $S$ will therefore be a $2N \times 2N$ matrix for the 
case of $N$ NM wires meeting at the junction with the SC region. Second, we 
will only study the conductances at energies much larger than $\De$, in 
contrast to the earlier sections where we only looked at the sub-gap 
conductances. The reason for considering energies much larger than 
$\De$ (but much smaller than the band width of the system) is that 
the analysis given below only works when the conductances are slowly varying 
functions of the energy; this will become clearer as we proceed.
We will take the 
interactions to be present only in the NM wires and derive RG equations which 
will tell us how the $S$-matrix evolves when we start at the length scale of 
the SC region and then increase the length scale to expand from that region 
into the NM wires.

In one dimension, the technique of bosonization is often used to
study systems with short-range density-density interactions~\cite{gogolin};
in this method a system of interacting fermions is mapped to a system of 
non-interacting bosons. This method has the advantage that interactions of 
any strength can be dealt with. However this method runs into difficulties in 
the presence of junctions of three or more wires and superconductivity for the 
following reasons. As we have seen, a junction is characterized by a matrix 
$\bf M$ which relates the electron fields on different wires in a linear
way. Since bosonization relates electron operators to exponentials of 
bosonic operators, a linear relation between fermionic fields in different 
wires translates, in general, to a non-linear relation between the bosonic 
fields; this makes it difficult to use bosonization. (However, there are some 
special forms of the junction matrix when bosonization can be applied. These
special cases correspond to the magnitudes of all the reflection or 
transmission amplitudes being equal to either zero or 1, namely, either
perfect reflection and no transmission, or no reflection and perfect 
transmission~\cite{lal}). Next, 
bosonization works best if the system is gapless and the energy-momentum 
dispersion is linear for both the fermionic and the bosonic theories which 
are related to each other. However superconductivity produces a gap 
proportional to the SC pairing $\De$, and the dispersion is not linear for 
energies of the same order as $\De$. So bosonization works only if we treat 
$\De$ as a perturbation (as was done in Ref.~\onlinecite{ganga} for example). 

We will therefore use a different approach which directly uses the fermionic 
language~\cite{yue,lal,das2,saha}. Unlike bosonization, this approach is 
useful only if the interaction strengths are weak in all the wires. However 
it has the advantage that it works for any form of the scattering
matrix which characterizes the junction.
The results obtained by this method and those obtained by bosonization
will of course match in the parameter regimes where both methods work, 
namely, when the interactions are weak and the junction scattering matrix 
has some special forms. We will now describe this method in detail.

We begin with a second quantized fermionic field $c(x)$ and the corresponding 
hole field $d(x)$, where $d(x)=c^\dg (x)$, for a single semi-infinite NM wire 
which goes from $x=0$ to $\infty$ and is 
connected to a SC region at $x=0$. At low temperatures, only low-energy 
processes are of interest and these only involve modes near the Fermi 
momenta $\pm k_F$; we will therefore consider only these modes. 
We therefore write the second-quantized field $c(x)$ as 
\beq c(x) ~=~ c_I(x) e^{-ik_F x} ~+~ c_O (x)e^{ik_F x}, \label{RG1} \eeq
where $c_I$ and $c_O$ denote the fields of incoming and outgoing electrons
respectively. We take these fields to be slowly varying on the length scale 
of $1/k_F$ as we have separated out the rapidly varying factors 
$e^{\pm ik_F x}$. Namely, the fields $c_I, ~c_O$ have momentum components 
$k$ such that $|k| \ll k_F$. We can then use a linear approximation for the 
dispersion relations of these fields so that $E=\pm v_F k$ for $c_I$ and
$c_O$ respectively, with $v_F$ being the Fermi velocity. 

Similarly, for the second-quantized field $d(x)$ we write 
\beq d(x)~=~ d_I (x) e^{ik_Fx} ~+~ d_O (x) e^{-ik_F x}. \label{RG2} \eeq
Note that the rapidly varying exponential terms multiplying $d_I$ and $d_O$ 
are the opposite of those multiplying $c_I$ and $c_O$. This is because 
destroying an electron is equivalent to creating a hole, so that $E_h=-E_e$ 
where $E_h$ and $E_e$ are the energies of the electron and hole respectively.
 
Let us now introduce a short-ranged density-density interaction between the 
electrons of the form
\beq H_{int} ~=~ \frac{1}{2} ~\int \int dx dy ~\rho(x) V(x-y) \rho(y). 
\label{int1} \eeq 
We assume that $V(x)$ is a real and even function of $x$. The density $\rho$ 
is a function of the second quantized fields given by $\rho(x) = c^\dg (x)c(x)
= d(x)d^\dg (x) = -d^\dg (x)d(x)$ (using the anticommutation property of the 
fermionic fields). Using Eq.~\eqref{RG1}, we obtain the expectation values
\bea <\rho(x)> &=& <c^\dg_I c_I> ~+~ <c^\dg_O c_O> \non \\
&& +~<c^\dg_I c_O> e^{2ik_F x} ~+~ <c^\dg_O c_I> e^{-2ik_F x}. \non \\
&& \label{int2} \eea

Next, we will assume that $V(x)$ is so short-ranged that $x$ and $y$, which 
appear as arguments of the density fields, can be set equal to each other 
except when the corresponding term in $H_{int}$ becomes zero. Using this 
assumption and the anticommutation relation between the fermionic fields, 
we get 
\beq H_{int} ~=~ g ~\int dx ~c^\dg_I c_I c^\dg_O c_O, \label{int31} \eeq 
where $g$ is related to the Fourier transform of $V(x)$ as $g = \tilde{V} (0)-
\tilde{V}(2k_F)$. From this expression it is clear that $g = 0$ if $V(x)$ is 
a $\de$-function. Hence $V(x)$ must have a finite range for
the interaction to have an effect. For each NM wire, the interaction is 
described by the single parameter $g$. The value of this parameter may be 
different for different wires which we will denote by $g_j$ on wire $j$. We 
define a dimensionless quantity $\al_j$ as 
\beq \al_j = \frac{g_j}{2 \pi v_F}, \label{int32} \eeq
where we assume $v_F$ to be the same in all the NM wires.
 
Let us briefly discuss how the interaction parameter appears in the formalism 
of bosonization. For spinless fermions, which is relevant here as we are 
studying $p$-wave SCs, the bosonic theory is characterized by two quantities, 
the velocity $v$ of the excitations and a dimensionless parameter $K$ (called 
the Luttinger parameter) which is a measure of the strength of the 
interactions between the fermions. These are related to $v_F$ and $\al$ as
\bea v &=& v_F (1-\al^2)^{1/2}, \non \\
K &=& \left( \frac{1-\al}{1+\al} \right)^{1/2}. \eea 
Note that $K=1$ if $\al = 0$ (non-interacting fermions), while $K<1$ ($K>1$) 
for $\al > 0$ ($\al < 0$), namely, repulsive (attractive) interactions. 
For weak interactions we get $v=v_F$ and $K=1-\al$ to first order in $\al$. 
We will do our RG analysis in the limit that $\al$ is small and positive in 
each wire.

As we have discussed earlier, it is generally difficult to bosonize a system 
with junctions. 
We will therefore use a different method which will enable us to derive 
RG equations directly for the scattering matrix of the junction. As we will 
see, this method only works up to first order in the interaction parameters 
$\al_j$. The basic idea of this method is the following. In the presence of 
non-zero reflection amplitudes $r_{jj}$ at the junction, the density of 
non-interacting fermions in the NM wire $j$ will have Friedel oscillations 
with wave number $2k_F$. When an interaction is turned on, an electron can 
scatter to an electron or a hole (Andreev reflection) from these oscillations 
with an amplitude which is proportional to the parameter $\al_j$. 
Ref.~\onlinecite{yue} used this idea to derive RG equations for an arbitrary 
$S$-matrix describing the junction of two semi-infinite wires. An RG analysis 
was then done for junctions of more than two wires, without superconductivity 
in Refs.~\onlinecite{lal} and \onlinecite{das2} and with a $s$-wave 
superconducting junction in Ref.~\onlinecite{saha}. We will carry out an
RG analysis for our system where several NM wires meet at a junction with 
a $p$-wave SC region. We expect the results to be much richer than those 
for a junction of NM wires when no superconductivity is present.


We will begin our analysis by deriving the form of the density oscillations 
in one particular NM wire close to the junction with the SC region. We will 
consider separately the cases of an electron and a hole coming in from a NM 
lead.

\vspace*{.6cm}
\noi {\bf 5.1 ~Processes related to an incoming electron}
\vspace*{.4cm}

An incoming electron can be either

\noi (i) normally reflected with amplitude $r_{ee}$, or

\noi (ii) Andreev reflected to a hole with amplitude $r_{he}$.
 
For momenta near $k_F$ we can write the wave functions for electrons and 
holes as 
\bea c_k (x) &=& c_I e^{-ik_F x} ~+~ c_O e^{ik_F x} \non \\
&=& e^{-i(k_F+k)x} ~+~ r_{ee} e^{i(k_F+k)x}, \non \\
d_k (x) &=& d_O e^{-ik_F x} ~=~ r_{he} e^{-i(k_F-k)x}, \label{cdx1} \eea 
where $|k|\ll k_F$. In the ground state of a non-interacting system, all 
the energy states below the Fermi energy $E_F$ are filled for electrons;
this corresponds to negative values of $k$ in Eq.~\eqref{cdx1}. Although
we are only interested in values of $k$ close to zero, it is mathematically 
convenient to take the range of $k$ to be $-\infty$ to $\infty$ even though 
the range is finite and given by the band width in real systems.
The expectation value of $\rho(x)$ in terms of $c_k (x)$ is given by 
\beq <\rho(x)> ~=~ \int_{-\infty}^0 \frac{dk}{2\pi} ~c^*_k c_k, 
\label{rhox} \eeq 
where we have taken the lower limit of $k$ to be $-\infty$.
We see that $<\rho>$ has a constant piece $\rho_0$ which can be eliminated 
by normal ordering. We are then left with
\beq <\rho(x)> ~-~\rho_0 ~=~ \frac{i}{4\pi x}(r^*_{ee}e^{-2ik_F x} ~-~ r_{ee}
e^{2ik_F x}). \label{Fo1} \eeq
It is clear that these terms arise entirely due to the interference between 
the incoming and reflected waves. Substituting Eq.~\eqref{Fo1} in 
Eq.~\eqref{int2} we see that $<c^\dg_I c_I+c^\dg_O c_O> = \rho_0$ is a 
constant, while 
\bea <c^\dg_O c_I> &=& \frac{ir^*_{ee}}{4\pi x}, \non \\
<c^\dg_I c_O> &=& <c^\dg_O c_I>^* ~=~ -~ \frac{ir_{ee}}{4\pi x}. \eea

An important point to note here is that there will also be a contribution to 
$c(x)$ and therefore to $<\rho(x)>$ from the waves which are transmitted 
from the other wires. As a wave transmitted to one wire from any other wire 
is incoherent with the incident and reflected waves of the first wire (we are 
assuming that waves incident from different NM leads are phase incoherent with 
respect to each other), there is no interference between these waves. Since 
the waves transmitted from the other wires only contribute to an outgoing wave 
in this wire, there is no interference and their contribution to $<\rho (x)>$
is independent of $x$. Hence, it can be absorbed in $\rho_0$. We conclude 
that the Friedel oscillations in Eq.~\eqref{Fo1} in a given wire arises 
only from the reflections within that wire. 

Now, in our system we have both normal and Andreev reflections. So there will 
be some non-zero expectation values for the operators which connect electrons 
and holes such as $c^\dg_I d_O$ and $d^\dg_O c_I$. Using Eq.~\eqref{cdx1}, we 
find that the expectation values of these operators are given by
\bea <c^\dg_I d_O> &=& \int_{-\infty}^0 \frac{dk}{2 \pi} ~r_{he} e^{i2kx} 
~=~ -~ \frac{ir_{he}}{4 \pi x}, \non \\
<d^\dg_O c_I>~&=&~ \frac{ir^*_{he}}{4 \pi x}. \label{cido} \eea
To evaluate the integral in the first equation in \eqref{cido}, we must 
introduce a factor like $e^{\ep k}$ which cuts off the contribution from
the lower limit $k \to - \infty$, and we then take the limit $\ep \to 0$.
Further, we have assumed that $r_{he}$ varies slowly with $k$ so 
that it is a reasonable approximation to take it outside the integral over 
$k$ in the first equation. This is the reason why our analysis only works at 
energies which lie far from the SC gap; for those energies the reflection
and transmission amplitudes are slowly varying functions of the energy.
In contrast to this, the sub-gap conductances have sharp peaks due to 
various sub-gap modes; hence the reflection and transmission amplitudes are 
not slowly varying functions of the energy if it lies inside the SC gap.
We note that a renormalization group study at energies within or close 
to the SC gap, where the reflection and transmission amplitudes vary rapidly, 
has been carried out in Ref.~\onlinecite{titov}.

Next, we derive the reflections of the electrons and holes from the Friedel 
oscillation by using the Hartree-Fock decomposition of the Hamiltonian in 
Eq.~\eqref{int31}. We have
\bea H_{int} &=& -g ~\int_0^{\infty} dx ~[<c^\dg_I c_O>c^\dg_O c_I ~+ 
<c^\dg_O c_I> c^\dg_I c_O \non \\
&& ~~~~~~+ <c^\dg_I c^\dg_O > c_I c_O ~+ <c_I c_O >c^\dg_I c^\dg_O ]. 
\label{int4} \eea

Using the expectation values of the various operators derived earlier and 
the identities $c_I c_O = c_I d_O^\dg = - d_O^\dg c_I$ and $c_I^\dg c_O^\dg 
= c_I^\dg d_O$, we obtain
\bea H_{int} &=& \frac{g}{4 \pi} ~\int_0^{\infty} \frac{dx}{x} ~
[i r_{ee} c^\dg_O c_I ~-~ i r^*_{ee} c^\dg_I c_O \non \\
&& ~~~~~~~~~~-~ i r_{he}d^\dg_O c_I ~+~ i r^*_{he}c^\dg_I d_O]. \eea

We can now derive the amplitude to go from an incoming wave to an outgoing 
wave under the action of $\exp(-i H_{int} t)$. We begin with an incoming 
electron with momentum $k$. Various processes can now occur.

\noi (i) An incoming electron with momentum $k$ can go to an outgoing 
electron with momentum $k^{\prime}$ under the action of $H_{int}$.
The corresponding amplitude is
\bea && -i ~\int \frac{dk^{\prime}}{2 \pi} ~2\pi \de (E_k - E_{k^{\prime}})
\non \\
&& \times ~<outgoing, e, k^{\prime}|H_{int}|incoming,e,k> \non \\
&& = ~\frac{g}{4 \pi v_F} ~\int_0^{\infty} \frac{dx}{x} ~r_{ee}~
e^{-2ikx}. \label{int5} \eea 
To obtain the above expression we have used Eq.~\eqref{int4}, the dispersion 
relation $E= v_F k$ (which implies $\de (E_k - E_{k^{\prime}})=
(1/ v_F)\de (k-k^{\prime})$), and the wave functions $e^{\pm{i(k_F+k)x}}$ 
of the outgoing and incoming electrons respectively.

To derive RG equations for quantities like $r_{ee}$ from Eq.~\eqref{int5}, we
will integrate $x$ over a small interval going from $L$ to $L + dL = L e^{dl}$.
Here $l$ is the logarithm of the length scale, and we can write $l = ln (L/a)$,
where $a$ is a short distance scale (which is the size of the superconducting 
region forming the junction) from which we will begin to integrate the RG 
equations. Eq.~\eqref{int5} then gives
\beq \frac{g r_{ee}}{4\pi v_F} ~dl ~=~ \frac{\al r_{ee}}{2} ~dl, 
\label{e1} \eeq 
where we have used Eq.~\eqref{int32}. 

\noi (ii) Similarly we find that the amplitude to go from an outgoing 
electron to an incoming electron is given by
\beq -~ \frac{\al r^*_{ee}}{2} ~dl. \label{e2} \eeq

\noi (iii) Due to the presence of the SC region, Andreev reflection can also 
occur, namely, an incoming electron can go to an outgoing hole under the 
action of $H_{int}$ as given in Eq.~\eqref{int4}. We can calculate the 
amplitude of this process in the same way as we did for normal reflection 
above. The amplitude to go from an incoming electron with momentum $k$ to an 
outgoing hole with momentum $k^{\prime}$ is found to be
\beq - ~\frac{\al r_{he}}{2} ~dl. \label{e3} \eeq

\noi(iv) The amplitude to go from an outgoing hole to an incoming electron is 
given by 
\beq \frac{\al r^*_{he}}{2} ~dl. \label{e4} \eeq
 
This completes the list of processes which can occur if we start with an 
incoming electron.
 
\vspace*{.6cm}
\noi {\bf 5.2 ~Processes related to an incoming hole}
\vspace*{.4cm}
 
All the processes we discussed in the previous section can be studied
if we start with an incoming hole. An incoming hole can be either

\noi (i) normally reflected to another hole with amplitude $r_{hh}$, or

\noi (ii) Andreev reflected to an electron with amplitude $r_{eh}$.

For momenta near $k_F$ we can write the hole and electron wave functions as 
\bea d_k (x) &=& d_I e^{ik_F x} ~+~ d_O e^{-ik_F x} \non \\
&=& e^{i(k_F-k)x} ~+~ r_{hh} e^{-i(k_F-k)x}, \non \\
c_k (x) &=& c_O e^{ik_F x} ~=~ r_{eh} e^{i(k_F+k)x}. \label{cdx2} \eea

For holes, we can use Eq.~\eqref{rhox} and the fact that $c^\dg c = d d^\dg
= -d^\dg d$ (plus a constant) to write the ground state expectation value of 
$\rho (x)$ as
\bea <\rho(x)> ~=~ - ~\int_{-\infty}^0 \frac{dk}{2\pi} ~d^*_k d_k. \eea
We can then find expectation values of various operators following a 
procedure similar to the previous section. We obtain
\bea <d^\dg_I d_O > &=& - \frac{i r_{hh}}{4 \pi x}, \non \\
<d^\dg_O d_I > &=& \frac{i r^*_{hh}}{4 \pi x}, \non \\
<c^\dg_O d_I > &=& \frac{i r^*_{eh}}{4 \pi x}, \non \\
<d^\dg_I c_O > &=& -\frac{i r_{eh}}{4 \pi x}. \eea
Using these expectation values we find the amplitudes of various processes.

\noi (i) Under the action of $H_{int}$, an incoming hole goes to an outgoing 
hole with an amplitude 
\beq \frac{\al r_{hh}}{2} ~dl. \label{h1} \eeq

\noi (ii) An outgoing hole goes to an incoming hole with an amplitude 
\beq -~ \frac{\al r^*_{hh}}{2} ~dl. \label{h2} \eeq

\noi (iii) An incoming hole goes to an outgoing electron with an amplitude 
\beq -~ \frac{\al r_{eh}}{2} dl. \label{h3} \eeq

\noi (iv) An outgoing electron goes to an incoming hole with an amplitude 
\beq \frac{\al r^*_{eh}}{2} dl. \label{h4} \eeq

\vspace*{.6cm}
\noi {\bf 5.3 ~RG equations for the reflection and transmission amplitudes}
\vspace*{.4cm}

The amplitudes of all the processes that we derived earlier can now be 
combined along with the $S$-matrix at a junction of several NM wires with a
SC region to calculate corrections to the $S$-matrix. We will calculate all 
the corrections to first order in the interaction parameters $\al_j$. 

We first consider corrections to the reflection amplitude $r_{ee,jj}$ on
wire $j$. To first order in $\al_j$, this gets contributions from the 
following processes. An incoming electron on wire $j$ can

\noi (i) become an outgoing electron on the same wire by scattering
from the Friedel oscillations in the density with the amplitude given in 
Eq.~\eqref{e1}.

\noi (ii) reflect from the junction with amplitude $r_{ee,jj}$ to 
become an outgoing electron, then become an incoming electron due to
the Friedel oscillations according to \eqref{e2}, and finally reflect
from the junction as an electron with amplitude $r_{ee,jj}$.

\noi (iii) reflect from the junction with amplitude $r_{ee,jj}$ to
become an outgoing electron, become an incoming hole due to 
Friedel oscillations according to \eqref{h4}, and then Andreev reflect
from the junction as an electron with amplitude $r_{eh,jj}$.

\noi (iv) Andreev reflect from the junction with amplitude $r_{he,jj}$
to become an outgoing hole, become an incoming electron 
due to Friedel oscillations according to \eqref{e4}, and then 
reflect from the junction as an electron with amplitude $r_{ee,jj}$.

\noi (v) become an outgoing hole by Andreev reflection from the 
junction with amplitude $r_{he,jj}$, become an incoming hole due to
Friedel oscillations according to \eqref{h2}, and then Andreev 
reflect from the junction as an electron with amplitude $r_{eh,jj}$.

\noi (vi) transmit through the junction to wire $i$ (with $i \ne j$) 
as an electron with amplitude $t_{ee,ij}$, turn from an outgoing 
electron to an incoming electron on wire $i$ due to Friedel oscillations
according to \eqref{e2}, and then transmit through the junction to wire 
$j$ as an electron with amplitude $t_{ee,ji}$.

\noi (vii) transmit through the junction into wire $i$ 
as an electron with amplitude $t_{ee,ij}$, turn from an outgoing 
electron to an incoming hole on wire $i$ according to \eqref{h4}, and
then transmit to wire $j$ as an electron with amplitude $t_{eh,ji}$.

\noi (viii) transmit through the junction into wire $i$ 
as a hole with amplitude $t_{he,ij}$, turn from an outgoing hole to an 
incoming electron on wire $j$ according to \eqref{e4}, and then transmit 
to wire $i$ as an electron with amplitude $t_{ee,ji}$.

\noi (ix) transmit through the junction into wire $i$ 
as a hole with amplitude $t_{he,ij}$, turn from an outgoing hole to an 
incoming hole on wire $j$ according to \eqref{h2}, and then transmit 
to wire $j$ as an electron with amplitude $t_{eh,ji}$.

Collecting all these terms, we find that the correction to $r_{ee,jj}$ is 
given by
\bea dr_{ee,jj} &=& B_{ee,jj} ~dl, \non \\
B_{ee,jj}&=& \frac{1}{2}[\al_j r_{ee,jj} - \al_j |r_{ee,jj}|^2 r_{ee,jj} 
+ \al_j |r_{eh,jj}|^2 r_{ee,jj} \non \\
&& + \al_j |r_{he,jj}|^2 r_{ee,jj} - \al_j r_{he,jj}r_{hh,jj}^* r_{eh,jj} 
\non \\
&& + \sum_{i \neq j} (-\al_i t_{ee,ji} r_{ee,ii}^* t_{ee,ij} + \al_i t_{ee,ji}
r_{eh,ii}^* t_{eh,ij} \non \\
&& ~~~~+ \al_i t_{he,ji} r_{he,ii}^* t_{ee,ij} - \al_i t_{he,ji}r_{hh,ii}^* 
t_{eh,ij} )]. \non \\
&& \label{r1} \eea

Similarly, the transmission amplitude $t_{ee,ji}$ from wire $i$ to wire 
$j$ can get corrections from the following processes. The incoming electron 
on wire $i$ can

\noi (i) get reflected from the junction as an electron with amplitude 
$r_{ee,ii}$, then become an incoming electron according to \eqref{e2}, 
and finally transmit to wire $j$ as an electron with amplitude $t_{ee,ji}$.

\noi (ii) reflect from the junction as an electron with amplitude 
$r_{ee,ii}$, become an incoming hole according to \eqref{h4}, and 
then transmit to wire $j$ as an electron with amplitude $t_{eh,ji}$.

\noi (iii) Andreev reflect from the junction as a hole with amplitude 
$r_{he,ii}$, become an incoming electron according to \eqref{e4}, and 
then transmit to wire $j$ as an electron with amplitude $t_{ee,ji}$.

\noi (iv) Andreev reflect from the junction as a hole with amplitude 
$r_{he,ii}$, become an incoming hole according to \eqref{h2}, and then 
transmit to wire $j$ as an electron with amplitude $t_{eh,ji}$.
 
\noi (v) transmit to wire $j$ as an electron with amplitude 
$t_{ee,ji}$, become an incoming electron on wire $j$ according to 
\eqref{e2}, and then reflect from the junction as an electron with 
amplitude $r_{ee,jj}$.
 
\noi (vi) transmit to wire $j$ as an electron with amplitude 
$t_{ee,ji}$, become an incoming hole on wire $j$ according to \eqref{h4},
and then reflect from the junction as an electron with amplitude $r_{eh,jj}$.

\noi (vii) transmit to wire $j$ as a hole with amplitude $t_{he,ji}$, 
become an incoming electron on wire $j$ according to \eqref{e4}, and then 
reflect from the junction as an electron with amplitude $r_{ee,jj}$.

\noi (viii) transmit to wire $j$ as a hole with amplitude $t_{he,ji}$, 
become an incoming hole on wire $j$ according to \eqref{h2}, and then 
reflect from the junction as an electron with amplitude $r_{eh,jj}$.

\noi (ix) transmit to wire $k$ (with $k \ne i, ~j$) as an electron with 
amplitude $t_{ee,ki}$, become an incoming electron according to 
\eqref{e2}, and then transmit to wire $j$ as an electron with amplitude 
$t_{ee,jk}$.

\noi (x) transmit to wire $k$ as an electron with amplitude $t_{ee,ki}$, 
become an incoming hole according to \eqref{h4}, and then transmit to 
wire $j$ with amplitude $t_{eh,jk}$.

\noi (xi) transmit to wire $k$ as a hole with amplitude $t_{he,ki}$, 
become an incoming electron according to \eqref{e4}, and then transmit 
to wire $j$ with amplitude $t_{ee,jk}$.

\noi (xii) transmit to wire $k$ as a hole with amplitude $t_{he,ki}$, 
become an incoming hole according to \eqref{h2}, and then transmit to wire 
$j$ with amplitude $t_{eh,jk}$.

Hence the total correction to $t_{ee,ji}$ is given by
\bea dt_{ee,ji} &=& B_{ee,ji} ~dl, \non \\
B_{ee,ji}&=& \frac{1}{2}[-\al_i t_{ee,ji}|r_{ee,ii}|^2+ \al_i 
t_{eh,ji} r_{ee,ii}r_{eh,ii}^* \non \\
&& +\al_i t_{ee,ji}|r_{he,ii}|^2 -\al_i t_{eh,ji} r_{he,ii}r_{hh,ii}^* \non \\
&& -\al_j |r_{ee,jj}|^2 t_{ee,ji} +\al_j |r_{eh,jj}|^2 t_{ee,ji} \non\\
&& +\al_j r_{ee,jj}r_{he,jj}^* t_{he,ji} - \al_j r_{eh,jj} r_{hh,jj}^* 
t_{he,ji} \non \\
&& +\sum_{k\neq i, j} (-\al_k t_{ee,jk} r_{ee,kk}^* t_{ee,ki} \non \\
&& +\al_k t_{eh,jk} r_{eh,kk}^* t_{ee,ki} + \al_k t_{ee,jk} r_{he,kk}^* 
t_{he,ki} \non \\
&& -\al_k t_{eh,jk} r_{hh,kk}^* t_{he,ki} )]. \label{t1} \eea

Similarly, we can find the corrections to all the other entries of the
$S$-matrix, namely, $r_{hh,jj}, ~r_{he,jj}, ~r_{eh,jj}, ~t_{hh,ji}, ~
t_{he,ji}$ and $t_{eh,ji}$. We now consider all these components of $S$ to be 
functions of a length scale $L$, where $L$ can vary all the way from a short 
distance scale $a$ to a large distance scale which may be either the length of
the NM wires or the thermal coherence length as discussed in 
Sect.~6.2. $L$ and $l$ are related as $l = ln (L/a)$, so that $l=0$ 
when $L=a$. Eqs.(\ref{r1}-\ref{t1}) then give us the RG equations
\bea \frac{dr_{ee,jj}}{dl} &=& B_{ee,jj}, \non \\
\frac{dt_{ee,ji}}{dl} &=& B_{ee,ji}. \label{RG} \eea
Eqs.~\eqref{RG} can be written in a more compact way. Given the matrix $S$ 
and the parameters $\al_j$, we define a matrix $F$ whose non-zero elements are
\bea F_{ee,jj} &=& \frac{1}{2} \al_j r_{ee,jj}, \non \\
F_{eh,jj} &=& -~ \frac{1}{2} \al_j r_{eh,jj}, \non \\
F_{he,jj} &=& -~ \frac{1}{2} \al_j r_{he,jj}, \non \\
F_{hh,jj} &=& \frac{1}{2} \al_j r_{hh,jj}. \label{fmat} \eea
We can then show that the RG equations for the different elements of 
$S$ (such as $r_{ee,jj}$ and $t_{ee,ji}$ given in Eq.~\eqref{RG}) can be
written compactly in the form of a matrix equation
\beq \frac{dS}{dl} ~=~ F - SF^\dg S. \label{RGflow} \eeq
This is the central result of this section. Note that these equations
are first order in the interaction parameters $\al_j$ since we have
only considered processes with {\it one} scattering from the Friedel
oscillations; hence the equations are valid only for weak interactions. 
(The interaction parameters $\al_j$ do not themselves flow under RG 
since we are considering a spinless system~\cite{solyom}).

We can verify from Eq.~\eqref{RGflow} that $S$ continues to remain unitary 
under the RG flows; it also remains symmetric if it begins with a symmetric 
form since $F$ is always symmetric. We also note that the form of 
Eq.~\eqref{RGflow} remains unchanged if $S$ is multiplied either from the 
left or from the right by a diagonal unitary matrix with entries of the form
\beq U_{jj} = e^{i\phi_j} , \label{unitary} \eeq
where the real numbers $\phi_j$ are independent of the length parameter $l$.
Hence the fixed points discussed below will also remain unchanged under 
such phase transformations. We will not distinguish between $S$-matrices 
which differ only by such phase transformations since experimentally we 
generally measure scattering probabilities rather than amplitudes.

\vspace*{.6cm}
\noi {\large {\bf 6 ~RG fixed points, stability analysis and conductance}}
\vspace*{.4cm}
 
\noi {\bf 6.1 ~RG fixed points}
\vspace*{.4cm}

We will now study the RG flows and find the fixed points and their stabilities.
For a system with a junction of $N$ wires, $S$ is a $2N \times 2N$ matrix 
which relates the $2N$-dimensional incoming and outgoing fields as $(c_O,~ 
d_O)^T = S (c_I,~ d_I)^T$. $S$ satisfies the following properties.
 
\noi (i) Unitarity: conservation of the particle current implies that 
$S^\dg S=I_{2N}$ where $I_{2N}$ is the $2N \times 2N$ identity matrix.

\noi (ii) Time reversal symmetry: under time reversal, we complex conjugate 
all numbers, change $t \to - t$, $c \to c^*$, and $d \to - d^*$. From 
Eqs.~\eqref{cdx1} and \eqref{cdx2}, we see that this transforms $c_I \to 
c_O^*$, $c_O \to c_I^*$, $d_I \to - d_O^*$, $d_O \to - d_I^*$. This implies 
that $S$ satisfies $S^\dg =\tau^z S^* \tau^z$, where \\
$\tau^z=\left( \begin{array}{cc} I_N & 0 \\ 0 & -I_N \end{array}\right)$,
and $I_N$ is the $N\times N$ identity matrix. We therefore get 
$S^T = \tau^z S \tau^z$.

If we want to find the fixed points analytically, we can use 
Eq.~\eqref{RGflow}, $dS/dl=F-SF^\dg S=0$, which implies that 
\beq SF^\dg ~=~ FS^\dg. \eeq
We can use this condition along with the two properties of $S$ mentioned 
above. After finding the fixed points of the RG equations, we can study their 
stabilities. To do this, we write a fixed point of the $S$-matrix as $S_0$, 
and a small deviation from this as $\ep S_1$, where $\ep$ is a small real 
parameter and $S_1$ is a matrix, namely, 
\beq S ~=~ S_0 ~+~ \ep S_1. \label{fp1} \eeq
For a given $S_0$, we can find the various flow `directions' $S_1$ such 
that Eq.~\eqref{RGflow} takes the form 
\beq \frac{d\ep}{dl}= \be \ep, \label{fp2} \eeq
where $\be$ is a real number. The solution of this equation is $\ep(l) = 
\exp (\be l) \ep (0)$ where $\ep(0)$ is given by the deviation of $S$ from 
$S_0$ at the short distance scale $a$. We see that $\be<0$ indicates that 
$S$ is stable against a perturbation in the direction of the corresponding 
$S_1$, while $\be > 0$ indicates an instability in the direction of $S_1$. 
The case $\be = 0$ describes a marginal direction; for instance, this
arises if we perturb a fixed point by a phase transformation as in 
Eq.~\eqref{unitary} which maintains it as a fixed point.

It turns out to be difficult to find all the fixed points and their 
stabilities analytically, particularly for the three-wire case. We will 
therefore study this problem numerically by starting with some randomly 
chosen matrix $S$, evolve it according to Eq.~\eqref{RGflow}, and see where 
it flows. However, we would like to begin with a matrix which satisfies the 
conditions $S^\dg S = I_{2N}$ (which implies non-linear constraints on the 
elements of $S$) and $S^T = \tau^z S \tau^z$. To generate such a matrix, let 
us write $S$ as $S= \exp (iA)$, where $A$ is also a $2N \times 2N$ matrix. 
Then the unitarity of $S$ implies $A^\dg =A$, and $S^T = \tau^z S \tau^z$ 
then implies that $A^T =\tau^z A \tau^z$. These provide linear 
relations between the elements of $A$ and are therefore easier to implement. 
(One can show that $N(2N+1)$ real parameters are required to specify such 
a matrix $A$). After randomly generating a matrix $A$ which satisfies these 
conditions, we take $S=\exp (iA)$ as the starting matrix for the RG flows
and see where it flows at large distance scales. 

\begin{figure}[h] 
\subfigure[]{\ig[width=3.0in,height=2.2in]{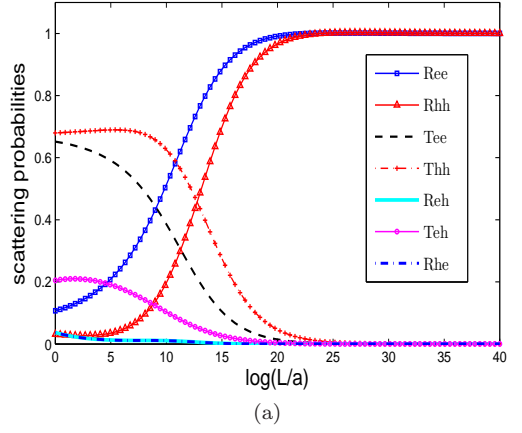}} \\
\subfigure[]{\ig[width=3.0in,height=2.2in]{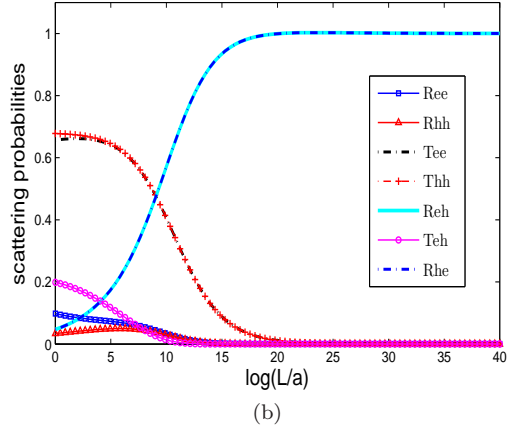}}
\caption[]{Typical RG flows of various scattering probabilities
for two wires meeting at a junction with a $p$-wave superconducting region.
Figure (a) shows the case with all $\al_j = 0.2$, while (b) shows the case
with all $\al_j = -0.2$.} \label{fig:2wirerg} \end{figure}

We first discuss a two-wire problem as a guide to the more complicated 
three-wire problem. Here we have two NM wires with a SC region present at 
their junction. Then $S$ is a $4 \times 4$ matrix of the form
\beq S= \left( \begin{array}{cccc}
r_{ee} & t_{ee} & r_{eh} & t_{eh} \\ 
t_{ee} & r_{ee} & t_{eh} & r_{eh}\\ 
r_{he} & t_{he} & r_{hh} & t_{hh} \\ 
t_{he} & r_{he} & t_{hh} & r_{hh} \end{array} \right). \label{smat2} \eeq
(The condition $S^T = \tau^z S \tau^z$ implies that $r_{he} = - r_{eh}$ and
$t_{he} = - t_{eh}$).
Here we have assumed for simplicity that all the scattering amplitudes are 
the same for wires 1 and 2, for instance, $r_{ee,11}=r_{ee,22}$ and so on;
these amplitudes can be complex in general. We will also assume that the
interactions parameters are equal on the two wires, $\al_1 = \al_2$.

We will first list the fixed points that we have found analytically for 
$S$-matrices of the form in Eq.~\eqref{smat2}. We will present the values of 
all the reflection and transmission amplitudes, with the understanding that 
two fixed points in which the amplitudes only differ by some phases as 
discussed in Eq.~\eqref{unitary} will not be considered as different fixed 
points. We have found the following fixed points which are similar to the ones
found for a junction with a $s$-wave superconductor in Ref.~\onlinecite{saha}.

\noi (i) $r_{ee} = r_{hh} = 1$, and all the other amplitudes are zero.

\noi (ii) $r_{eh} = - r_{he} = 1$, and all the other amplitudes are zero.

\noi (iii) $t_{ee} = t_{hh} = 1$, and all the other amplitudes are zero.

\noi (iv) $t_{eh} = - t_{he} = 1$, and all the other amplitudes are zero.

\noi (v) $r_{ee} = t_{hh} = r_{he} = t_{he} = 1/2$, and $t_{ee} = r_{hh} =
r_{eh} = t_{eh} = -1/2$.

\noi (vi) $t_{ee} = t_{hh} = \cos \ta$, $t_{eh} = - t_{he} = \sin \ta$, and
all the other amplitudes are zero. Here $\ta$ can be any real number from 0 
to $2\pi$. We thus have a continuous family of fixed points labeled by $\ta$.
This is in contrast to the fixed points (i), (ii) and (v) given above which 
are all discrete fixed points. We note that the fixed points in (iii) and (iv)
are special cases of (vi) corresponding to $\ta = 0$ and $\pi/2$ respectively.

A simple way of finding fixed points analytically is to set all the reflection 
amplitudes equal to zero, $r_{ee,jj}=r_{hh,jj}= r_{eh,jj}= r_{he,jj}= 0$. Then
the matrix $F$ in Eq.~\eqref{fmat} vanishes, and Eq.~\eqref{RGflow} 
straightaway gives $dS/dl=0$. All the other amplitudes in $S$ are then 
constrained by unitarity. This is how the family of fixed points in (vi) 
was found.

We will now numerically study the fixed points and their stabilities for
Eq.~\eqref{smat2}. We begin by randomly generating a matrix $A$ satisfying 
the conditions $A^\dg = A$ and $A^T = \tau^z A \tau^z$. 
Given an $A$, we construct $S=\exp (iA)$, and then use Eq.~\eqref{RGflow} to 
evolve $S$. In Figs.~\ref{fig:2wirerg} (a) and (b), we show the typical 
RG flows of a number of reflection and transmission probabilities, such as
$R_{ee} = |r_{ee}|^2$, $R_{hh} = |r_{hh}|^2$, etc. We find numerically 
that for a general starting point, $S$ always flows to the same fixed point
which is described below. (By a general starting point, we mean that we do 
not start precisely at a special set of matrices which constitute some other 
fixed points which are not completely stable). For the case where $\al_1 = 
\al_2 = 0.2$ (i.e., the interactions are repulsive), 
the plots in Fig.~\ref{fig:2wirerg} (a) show that $S$ flows to a fixed point 
where $R_{ee}=R_{hh}=1$, and all the other probabilities are zero. If we start
the RG evolution exactly at this point, $S$ does not flow at all and remains 
at that point. Hence this is the fixed point in the case of repulsive 
interactions. This is a completely stable fixed point because the RG flows 
always approach this fixed point independent of the $S$ that we start from; 
further, if we deviate a little bit in any direction from this fixed point, 
the RG flows take us back to the fixed point. 

Interestingly, we find even if none of the symmetries are present, namely, 
$\al_1 \ne \al_2$ (but both are positive), there is no time reversal symmetry 
($S^T \ne \tau^z S \tau^z$), and the starting $S$-matrix is not symmetric 
between the two wires, $S$ always flows to a completely stable fixed point 
which is symmetric. More carefully speaking, one finds at the fixed point 
that although the phases of $r_{ee,jj}$ and $r_{hh,jj}$ are generally not 
equal to each other (the phases depend on the starting value of $S$ and the 
values of $\al_j$), all their magnitudes are equal to 1. We thus have a 
{\it symmetry restoration} at the fixed point.
 
For the opposite case in which $\al_1 = \al_2 = -0.2$ (i.e., attractive 
interactions), the plots in Fig.~\ref{fig:2wirerg} (b) 
show that $S$ generally flows to a different fixed point where $R_{eh}=
R_{he}=1$, and all the other probabilities are zero. This is the completely 
stable fixed point in this case as we always reach this point no matter 
where we start (unless again we start precisely at some special set of 
matrices which are fixed points which are not completely stable), and
small deviations from this fixed point in any direction also flow to zero. 
Once again, we find that the phases of $r_{eh,jj}$ and $r_{he,jj}$ are 
generally not equal to each other but all their magnitudes are equal to 1 
at the fixed point, even if there is no symmetry in the starting value of 
$S$ and of the $\al_j$ (provided that they are both negative).

We now observe that since Eq.~\eqref{RGflow} is linear in the $\al_j$, 
the equation depends only on the combination $\al_j l$. 
Hence the direction of the RG flows for $\al_j > 0$ and
$l \to \infty$ is just the opposite of the flows for $\al_j < 0$ and
$l \to \infty$. We therefore conclude that the stable fixed points for 
attractive interactions are exactly the same as the unstable fixed points 
for repulsive interactions. So the fixed point with $R_{eh} = R_{he}=1$ 
is the completely unstable fixed point for repulsive interactions, i.e.,
if we start slightly away from this point in any direction, the RG flows
will always take us further away from the point.
 
We note that this numerical way of finding fixed points cannot 
detect fixed points which are partially stable and partially unstable, 
namely, stable in some directions and unstable in other directions.
If we start near such a fixed point, we will generally always flow 
away from it regardless of whether we take all the $\al_j$ to be 
positive or negative.


Finally, let us discuss the RG fixed points and their stabilities for the 
three-wire problem. We again assume for simplicity that there is complete 
symmetry between the three wires, both in the $S$-matrix and the interactions 
strengths. $S$ is therefore a $6 \times 6$ matrix of the form
\beq S= \left( \begin{array}{cccccc}
r_{ee} & t_{ee} & t_{ee} & r_{eh} & t_{eh} & t_{eh} \\
t_{ee} & r_{ee} & t_{ee} & t_{eh} & r_{eh} & t_{eh} \\
t_{ee} & t_{ee} & r_{ee} & t_{eh} & t_{eh} & r_{eh} \\
r_{he} & t_{he} & t_{he} & r_{hh} & t_{hh} & t_{hh} \\
t_{he} & r_{he} & t_{he} & t_{hh} & r_{hh} & t_{hh} \\
t_{he} & t_{he} & r_{he} & t_{hh} & t_{hh} & r_{hh} \end{array} \right). 
\label{smat3} \eeq

We now present the results we find numerically, by starting with a 
randomly chosen matrix $S=\exp (iA)$ which has all the desired symmetries and 
then evolving it using Eq.~\eqref{RGflow}. The typical RG flows of the various 
scattering probabilities are very similar to the ones shown for a two-wire 
system in Figs.~\ref{fig:2wirerg} (a) and (b) for all the $\al_j$ equal to 
$0.2$ (repulsive) and $-0.2$ (attractive interactions) respectively. We see 
that the completely stable fixed point for repulsive interactions is again 
given by $R_{ee}=R_{hh}=1$, and all the other probabilities are zero. As we 
discussed above, the stable fixed points for attractive interactions, i.e., 
$\al_j<0$, are the unstable fixed points for repulsive interactions. Using 
this property we find that the completely stable fixed point for attractive 
interactions, and therefore the completely unstable fixed point for repulsive 
interactions, is given by $R_{eh}=R_{he}=1$, and all the other probabilities 
are zero. All these statements remain true even if the $\al_j$ are not equal 
to each other (although they must all have the same sign) and even if the 
starting value of $S$ has no symmetries. Thus the completely stable and 
completely unstable fixed points are similar for the two-wire and three-wire 
systems. 

Before ending this section, we would like to mention the work done in
Ref.~\onlinecite{affleck} on a junction of a superconducting wire and two 
non-superconducting wires where there are interactions. A non-trivial stable 
fixed point was found there which has perfect normal reflection for 
one linear combination of the electron fields in the two wires and
perfect Andreev reflection for the other linear combination. This fixed
point occurs because Ref.~\onlinecite{affleck} considers an interaction
between the wires which mixes the two electron fields. We have not
considered such inter-wire interactions in our model and therefore
do not find such a non-trivial fixed point.

\vspace*{.6cm}
\noi {\bf 6.2 ~Conductances under RG flows}
\vspace*{.4cm}

We will now use our understanding of the RG flows to study the conductances 
of a three-wire system as functions of physical parameters such as the wire 
lengths and the temperature, for the case of repulsive interactions. In 
particular, we will study how the conductances scale with various lengths 
when we approach the completely stable fixed point. 

Using the procedure described in Eqs.~\eqref{fp1} and \eqref{fp2}, we first 
find the values of $\be$ for different flow directions given by the 
perturbation $S_1$ around a fixed point $S_0$. The stable fixed point that
we are interested in has only the $r_{ee,jj}$'s and $r_{hh,jj}$'s (namely,
the diagonal elements of $S_0$) being unimodular numbers and all the other 
elements being zero. For a flow direction $S_1$ in which only the phases of 
the diagonal elements are changed, we find that $d\ep /dl= 0$. This implies 
that $\be=0$ and the RG flow is marginal. This is expected since the RG fixed 
points are invariant under multiplication by a diagonal unitary matrix as 
discussed in Eq.~\eqref{unitary}; hence there is no RG flow in those 
directions. Next, we look at the RG flows when only the $r_{eh,jj}$'s and 
$r_{he,jj}$'s are perturbed from zero. For this perturbation, we find that 
$d\ep /dl= - 2\al \ep$; hence $\be=-2\al$, and the RG flow in this direction 
is irrelevant. We choose $S_1$ where either (i) only 
the $t_{ee,ij}$'s and $t_{hh,ij}$'s are non-zero, or (ii) only the 
$t_{eh,ij}$'s and $t_{he,ji}$'s are non-zero. For both these cases we find
that $d\ep /dl= -\al \ep$, which means that $\be=-\al$ and the RG flows
in these directions is also irrelevant. The RG flows near the fixed point 
$S_0$ are therefore either marginal or irrelevant in all directions.

Now we will discuss how the Cooper conductance $G_C$ and the normal 
conductances $G_{N2}$ and $G_{N3}$ scale under the RG flows (we are assuming
that an electron is incident from the NM lead 1). Physically there are 
three length scales in the problem and we have to stop the RG flows when we 
reach the smallest of the three scales. One length scale is $\eta = \hbar 
v_F /\De$ which is associated with the SC gap, another is the wire length 
$L_w$ (we will take the lengths of all the three wires to be of the order of 
$L_w$), and the third scale is the thermal coherence length $L_T= \hbar v_F / 
(k_B T)$ if the system is at a temperature $T$~\cite{lal,das2}. (We assume
that all these length scales are much larger than the short distance scale 
$a$). We will now consider different regimes of these length scales.

\noi (i) We first consider the case in which $L_w$ is finite and smaller than
$\eta$, while $T \longrightarrow 0$ so that $L_T \longrightarrow \infty$. Then 
the length scale where the RG flows must be stopped is $L_w$. For a general 
perturbation around the stable fixed point, we have $|r_{he}|^2 \sim 
e^{-4\al l}$, $|t_{he}|^2\sim e^{-2\al l}$, and $|t_{ee}|^2\sim e^{-2\al l}$, 
where $l=ln(L_w/a)$. We then find that $G_C = 2 (|r_{he}|^2 + 2 |t_{he}|^2)$
scales as
\bea G_C &\sim& c_1 e^{-4 \al l} ~+~ c_2 e^{-2 \al l} \non \\
&\sim& \frac{c_1}{(L_w/a)^{4\al}} ~+~ \frac{c_2}{(L_w/a)^{2\al}}, 
\label{gc1} \eea
where $c_1, ~c_2$ are some constants which depend on how far from the
stable fixed point we are at the length scale $a$ where the RG flows begin.
If $L_w \gg a$, the second term in Eq.~\eqref{gc1} dominates over the
first term. We therefore obtain
\beq G_C ~\sim~ \frac{1}{(L_w/a)^{2\al}}. \label{gc2} \eeq
Similarly we find that $G_{N2}, ~G_{N3} \sim |t_{ee}|^2 - |t_{he}|^2$ scale as
\beq G_{N2}, ~G_{N3} ~\sim~ \frac{1}{(L_w/a)^{2\al}}. \label{gn23} \eeq

\noi (ii) Next we consider the case in which $T$ is finite such that $L_T
=\hbar v_F/(k_B T)$ is smaller than $\eta$, and $L_w \longrightarrow \infty$.
Then the RG flows stop at the length scale $L_T$. Hence the elements of $S$ 
must be evaluated at $l=ln(L_T/a) = ln( v_F/(k_B T a))$. Hence
\beq G_C ~\sim~ c_1 (k_B T a/\hbar v_F)^{4\al} ~+~ c_2 (k_B T a/\hbar 
v_F)^{2\al}. \eeq
In the limit $k_B T a/\hbar v_F \ll 1$, the second term dominates over the 
first and we get
\beq G_C ~\sim~ (k_B T a/\hbar v_F)^{2\al}. \eeq
Similarly, we find 
\beq G_{N2}, ~G_{N3} ~\sim~ (k_B T a/\hbar v_F)^{2\al}. \eeq

\noi (iii) Finally we consider the case where $\eta = \hbar v_F /\De$ is 
smaller than both $L_w$ and $L_T$. Then the RG flows must be stopped at the 
length scale $\eta$ since our RG equations were derived under the condition 
that the energy scale is much larger then $\De$. We then obtain expressions 
for the conductances which are similar to Eqs.~\eqref{gc2} and \eqref{gn23} 
but with $L_W$ replaced by $\eta$.

To conclude, we see that the conductances $G_C$ and $G_{Nj}$ approach zero
as powers of the smallest length of the system which may be $L_w$, $L_T$ or
$\eta$.
The exponents of the power laws can give an
estimate of the strength of the interactions in the wires.

\vspace*{.6cm}
\noi {\large {\bf 7 ~Experimental realization of systems with different signs 
of $\De_j$}}
\vspace*{.4cm}

We will now discuss how it may be possible to experimentally realize a
system of three SC wires with the same or different signs of the $p$-wave 
pairings $\De_j$. We consider the system studied in Ref. \onlinecite{ganga}. 
This consists of a wire with a Rashba spin-orbit coupling of the form $\pm 
\al_R p_r \si^x$, where $p_r$ is the momentum along the wire and $\si^x$ is a 
Pauli spin matrix. This form can arise as follows. Let us take the coordinate
in the wire to increase along an arbitrary direction $\hat r$ lying in the 
$x-y$ plane. If the Rashba term is
$\al_R {\hat n} \cdot {\vec \si} \times {\vec p}$, and $\hat n$ points in the
$\hat z$ direction, then the Rashba term will be $\al_R p_r \si^x$ if
${\hat r}={\hat y}$ and $- \al_R p_r \si^x$ if ${\hat r}=-{\hat y}$.

Next, we place the wire in a magnetic field in the $\hat z$ direction which 
is perpendicular to the Rashba term (and has a Zeeman coupling $\De_Z$ to 
the spin of the electrons) and in proximity to a bulk $s$-wave SC 
with pairing $\De_S$. The complete Hamiltonian is~\cite{ganga}
\bea H &=& \int dr ~\Psi^\dg_\al \left[ (\frac{p_r^2}{2m} - \mu)\de_{\al \be}
\pm \al_R p_r \si^x_{\al \be} - \De_Z \si^z_{\al \be} \right] \Psi_\be \non \\
&& + \frac{i}{2} \int dr ~[\De_S \Psi^\dg_\al \si^y_{\al \be} \Psi^\dg_\be +
H.c.], \label{rashba} \eea
where $\Psi_\al$ is the annihilation operator for an electron with spin $\al$,
and the $\pm$ sign of the Rashba term depends on whether ${\hat r} = \pm 
{\hat y}$. Ref. ~\onlinecite{ganga} then shows that for a
certain range of the parameters, this system is equivalent to a spinless
$p$-wave SC of the form that we have studied in this paper, with the $p$-wave
pairing term being given by $-i (\De/k_F) (c^\dg \pa_x d + d^\dg \pa_x c)$ 
(see Eq.~\eqref{Ham}), where
\beq \De ~=~ \pm ~\frac{\al_R k_F \De_S}{\De_Z}. \label{eff} \eeq

\begin{figure}[h] 
\subfigure[]{\ig[width=1.1in]{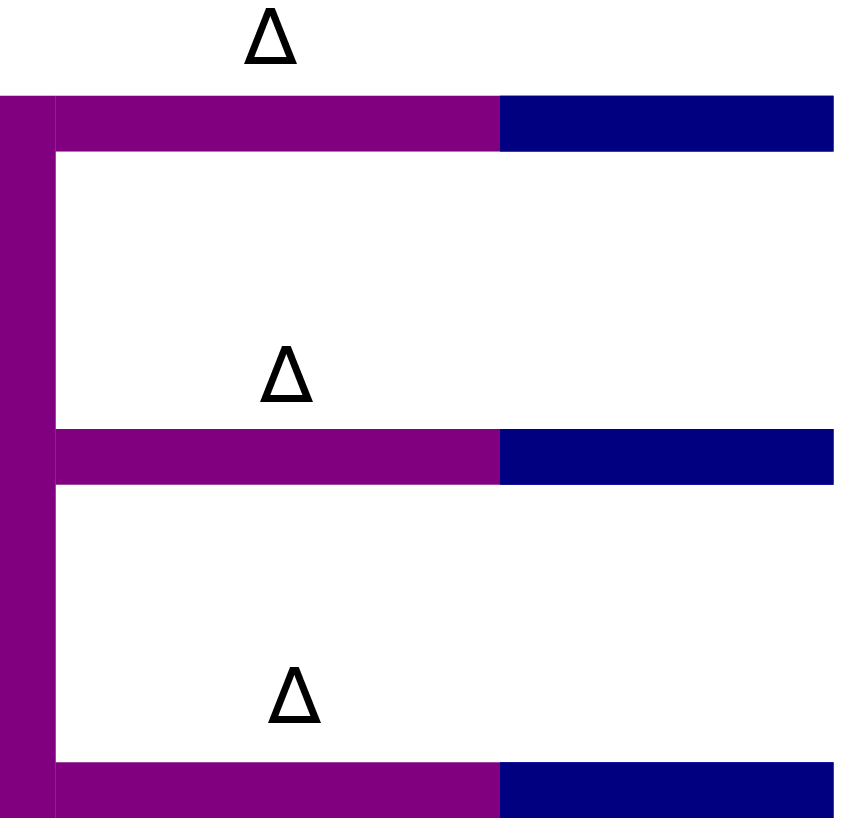}} \\
\vspace*{.3cm}
\subfigure[]{\ig[width=2.0in]{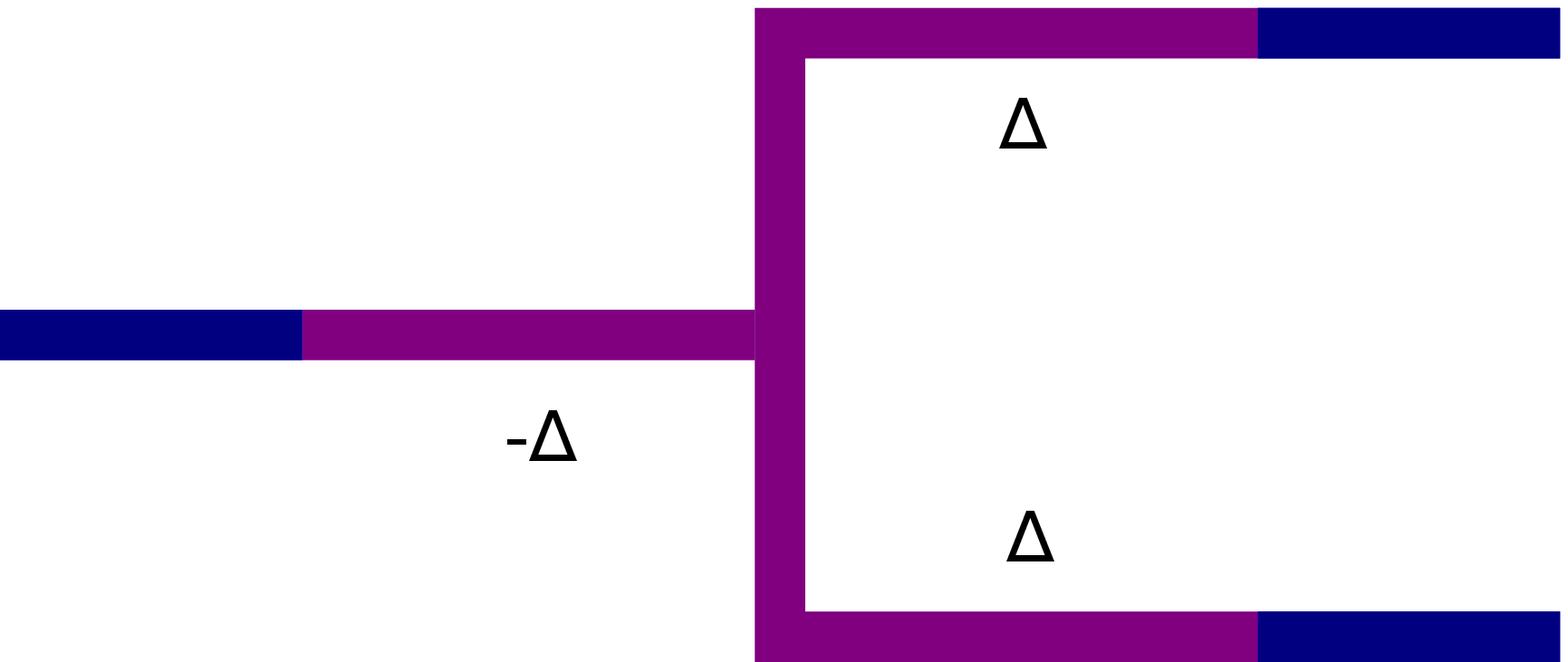}}
\caption[]{Three SC wires with (a) the sign of $\De_j$ being 
the same in all the wires, (b) the sign of $\De_j$ being different in the 
wire on the left compared to the other two wires. The SC wires are shown in 
a lighter shade while the NM leads are shown in a darker shade.} 
\label{fig:exp} \end{figure}

Now consider a case in which the coordinates are given by ${\hat r} =
{\hat y}$ in all the three wires; see Fig.~\ref{fig:exp} (a). (The region 
where the three SC wires meet is an extended vertical region on the left 
side of the figure). Then the Rashba term and 
hence $\De_j$ will have the same sign in all the SC wires. On the other hand, 
suppose that one of the SC wires runs in a direction which is opposite to the 
other two wires as shown in Fig.~\ref{fig:exp} (b). (The three SC wires now 
meet along an extended vertical region in the middle of the figure). 
Now ${\hat r} = -{\hat y}$ in that wire while ${\hat r} = {\hat y}$ in the 
other two wires. Then Eq.~\eqref{eff} shows that $\De_j$ will have one sign 
in that wire and the opposite sign in the other two wires.
 
[We would like to note that if the angle between the wires is different from
zero or $\pi$, the situation will be more complicated because the Rashba term
$\al_R {\hat n} \cdot {\vec \si} \times {\vec p}$ will no longer be
proportional to the same $\vec \si$ matrix in the different wires.
Hence the effective $p$-wave pairings in the different wires will not
be related simply by sign changes in $\De_j$].

It is clear that the vertical region where the three SC wires meet is likely 
to cause some scattering of the electrons; we have modeled this scattering in 
the earlier sections using the matrix $\bf M$. Finally, the ends of the SC 
are connected to NM leads through tunnel barriers. As discussed in 
Sect.~2, these barriers can be characterized by their strength $\la$.

In Sects. 2 - 4, we discussed a conductance $G_C$ in which pairs of electrons 
can appear in (or disappear from) the SC regions. At a microscopic level 
we can understand these processes as occurring due to a Cooper pair going from
the bulk $s$-wave SC to one of the $p$-wave SC wires (or vice versa). Finally 
we assume that the $s$-wave SC is grounded, and the three NM leads and the 
$s$-wave SC form a closed electrical circuit so that we can measure the 
conductances $G_C$ and $G_{Nj}$.

\vspace*{.6cm}
\noi {\large {\bf 8 ~Conclusions}}
\vspace*{.4cm}

In this paper, we have studied the conductances of a 
system consisting of three $p$-wave superconductors, each connected to 
a normal metal lead (these are labeled as SC1, SC2, SC3 and NM1, NM2, NM3
respectively) in two different regimes of the energy. 
In Sects. 2 - 4, we have studied the Majorana and related
sub-gap modes and their effects on the sub-gap conductances. Here
we have considered two cases: (i) when the $p$-wave pair potentials $\De_j$ 
have the same sign in all the SCs, and (ii) when one of the $\De_j$ has 
a different sign from the other two. In Sects. 5 - 6,
we have studied the effect of interactions between the electrons on the 
conductances at energies lying far from the SC gap.

To study the sub-gap modes,
we have used a continuum model and current conservation to derive the 
boundary conditions at the junctions between the NMs and the SCs. Then 
we have found the boundary condition at the junction of the three SCs; this 
condition is encoded by a Hermitian matrix $\bf M$. Using these conditions, we 
have numerically studied two conductances, the Cooper pair conductance 
$G_C$ (from NM1 to the SCs) and the normal conductance $G_{N2}$ (from NM1 
to NM2) when we send in an electron from NM1 with an energy $E$; we have 
taken $E$ to lie inside the superconducting gap so as to specifically 
probe the sub-gap modes which reduce to zero energy Majorana modes in
the limit of long wire lengths.

We have looked at three different regimes of the lengths of the SCs with 
respect to $\eta$ which is the length scale associated with the SC gap. We find 
a rich pattern of the conductances as functions of the SC lengths and $E$. 

We first consider the case when all the $\De_j$'s have the same sign. In the 
length regime $L \simeq \eta$ we find six sub-gap modes; these are not at 
zero energy because they hybridize with each other due to the finite 
lengths of the SCs. Among the six modes, three sit near the SC-NM 
junctions, and the other three sit at the junction of the three SCs. All these 
modes have a significant effect on the conductances and the latter shows 
peaks exactly at the energies of the sub-gap modes. 

For the length regime $L \simeq 2 \eta$ the hybridization among the sub-gap 
modes is less than in the previous case; hence the energy splitting of 
the sub-gap modes is also less. Even if we cannot clearly see six separate 
sub-gap modes, we can still observe several modes at non-zero energies.

In the third regime, when $L \gg \eta$, we find that the sub-gap modes are 
almost decoupled from each other and hence lie at zero energy. The Cooper 
conductance $G_C$ almost approaches its highest value of $2e^2 /h$. 

We compare our conductance results with the sub-gap energies that we get for 
a box made of three SC wires with hard walls (namely, without the NM leads). 
We generate plots for the energies of the sub-gap modes in all the three 
length regimes as discussed earlier. In the first length regime, instead of 
six sub-gap modes we see only four clearly; we believe this is due to the 
fact that two of the modes are very close to $E/ \De = \pm 1$ and hence lie 
beyond our resolution. In the other two regimes, the plots of the energies
of sub-gap modes almost exactly match with the conductance plots for the full 
system as we can see in Fig.~\ref{fig:3sc2}.

In obtaining all these results, we have made a particular choice of the 
matrix $\bf M$ which defines the boundary condition at the junction of 
three SCs, namely, we have taken all the diagonal and off-diagonal elements 
of $\bf M$ to be equal to 1. If we take the elements to be different, the 
results may be somewhat different from what we get.

With the same choice of $\bf M$, we have calculated the conductances 
taking the $\De$ in SC1 to have a different sign compared to the $\De$'s in 
SC2 and SC3. The results are significantly different from the case of 
all $\De_j$'s having the same sign. In the length regime $L \simeq \eta$ 
we find four sub-gap modes at different energies, instead of six. Now there 
is only one sub-gap mode sitting at the junction of three SCs. 
As we increase $L/\eta$, the energy splitting between the sub-gap modes 
decreases; when $L \gg \eta$ we find that all the modes lie at zero energy.
Once again we have compared our conductance results with the energies of the
sub-gap modes in a SC box with the $\De$ in SC1 having an opposite sign to 
SC2 and SC3. The results match very well as we can see in Fig.~\ref{fig:3sc3}.

We have then presented some analytical and symmetry arguments to explain the 
presence of multiple zero energy Majorana modes at a junction of three long 
SCs. Using the boundary condition at the junction we find that the number of 
independent variables in the problem directly corresponds to the number of 
Majorana modes at the junction. We have also used an ``effective time 
reversal symmetry" to argue that the number of Majorana modes at a junction 
of three Kitaev chains can be either one or three.

Next, we have studied the effect of interactions between the electrons 
in order to understand the conductances at energies far from the SC gap. 
(We emphasize that the energy range we have considered is different from
earlier work where RG equations have been studied at energies within or
close to the SC gap, where the various reflection and transmission amplitudes 
vary rapidly with the energy~\cite{fidkowski2,titov}). We have 
derived the RG equations for a general scattering matrix $S$ which governs a 
system of several NM wires which meet at a junction with a SC region. The RG 
flows of the elements of $S$ are entirely a result of the interactions in the 
NM wires; there is no flow if all the interaction parameters $\al_j$ are 
equal to zero. We have found stable and unstable fixed points for 
two-wire and three-wire junctions when the interactions are repulsive, i.e.,
$\al_j>0$. A completely stable fixed point is one where the RG flows take 
us back to the fixed point for a small deviation from it in any direction. 
We find that at the completely stable fixed point, the magnitudes of all the 
normal reflection probabilities are equal to 1, while at the completely
unstable fixed points all the Andreev reflection probabilities are equal to 1.
We have discussed the consequences of the RG flows on the conductances of the
system, assuming all the interaction parameters to be equal to $\al$. A 
stability analysis near the stable fixed point shows that when the 
wire lengths $L_w$ are finite and the temperature $T \longrightarrow 0$, 
the Cooper conductance $G_C$ and the normal conductances $G_{N2}, ~G_{N3}$
all scale as $1/L_w^{2\al}$. In the other limit where $T$ is finite (but
small compared to the band width of the system), and $L_w \longrightarrow 
\infty$, $G_C$, $G_{N2}$ and $G_{N3}$ scale as $T^{2\al}$. In general,
for an electron incident from NM lead $j$, the conductance measured
on NM lead $j$ will scale as $1/L_w^{2\al}$ or $T^{2\al}$ depending on
which length scale is smaller. Thus a measurement of the conductances
can provide valuable information about the strength of the interactions
between the electrons.

We can summarize our most important results as follows.

\noi (i) We have shown that a system of three $p$-wave SC wires with 
NM leads has multiple sub-gap modes and all these modes contribute to peaks 
in the conductance; the positions of the peaks exactly match the 
energies of the sub-gap modes. We thus have a novel system with a large number
of sub-gap conductance peaks which vary with the applied bias and wire 
lengths in an interesting way.

\noi (ii) To study the conductances far from the SC gap, we have used an RG 
method which
directly uses the fermionic language to look at the effect of interactions
on the scattering matrix. (This is in contrast to earlier RG studies of 
$p$-wave superconductors which used bosonization and studied the RG flows
of parameters in the bosonic Hamiltonian rather than the scattering matrix).
We have shown that interactions make the scattering matrix flow to
certain stable fixed points at large wire lengths and low temperatures.
The fixed points are symmetric under permutations of the wires and under 
time reversal even if the system does not have these symmetries at the 
microscopic length scale. Near the stable fixed point, we have shown
that the conductances scale as a power of the wire length or the temperature,
and the power can give us an estimate of the strength of the interactions.

Putting the above results together, we get a complete picture of the 
conductances of a system of three $p$-wave SC wires both inside and 
far outside the SC gap.

Finally, we have discussed how our model can be experimentally implemented
by taking three wires with Rashba spin-orbit coupling, applying a Zeeman field 
perpendicular to the direction of the Rashba field, and placing the system
in proximity to an $s$-wave superconductor. In a particular range of
parameters, each wire effectively becomes a $p$-wave SC. The cases where 
the three SC wires have the same sign of the $p$-wave pairing or one of them 
has the opposite sign of the pairing can both be implemented, simply by 
rearranging the orientation of one of the wires. As we have shown in the 
earlier sections, the number of sub-gap modes is different in the two cases, 
and this leads to a significant difference in the pattern of conductance peaks
as a function of the energy and wire lengths. In future, we can generalize
our studies of a single three-wire junction to more complicated systems such 
as networks of wires forming a lattice~\cite{alicea1,zhou}, where each vertex 
of the lattice can host one or more sub-gap modes.

\section*{Acknowledgments}

We thank S. Das, J. N. Eckstein, S. Rao and A. Soori for stimulating 
discussions. For financial support, M.T. thanks CSIR, India and D.S. thanks 
DST, India for Project No. SR/S2/JCB-44/2010.

\end{document}